\pdfoutput=1
\documentclass[a4paper,twocolumn,11pt,unpublished,nopdfoutputerror,allowtoday]{quantumarticle}

\usepackage{amsmath,amssymb}
\usepackage{bm}
\usepackage{amsthm}
\usepackage{subfigure,overpic}
\usepackage{mathrsfs}
\usepackage{multirow}
\usepackage{graphicx}
\usepackage{physics}
\usepackage{cancel}
\usepackage[numbers,sort&compress]{natbib}
\usepackage[colorlinks=true,linkcolor=blue,citecolor=blue,urlcolor=blue]{hyperref}

\def\dbar{{\mathchar'26\mkern-12mu d}}

\begin{document}

\title{A thermodynamic uncertainty relation for (hybrid) N--S coherent conductors}

\author{Sergi Vidal}
\affiliation{Instituto de F\'isica Interdisciplinar y Sistemas Complejos,
  IFISC (CSIC-UIB), Campus Universitat de les Illes Balears,
  E-07122 Palma de Mallorca, Spain}

\author{Rosa L\'opez}
\affiliation{Instituto de F\'isica Interdisciplinar y Sistemas Complejos,
  IFISC (CSIC-UIB), Campus Universitat de les Illes Balears,
  E-07122 Palma de Mallorca, Spain}

\date{\today}

\begin{abstract}
Thermodynamic uncertainty relations establish fundamental bounds between current fluctuations and entropy production in nonequilibrium systems. In hybrid normal–superconducting conductors, transport is governed by the coexistence of quasiparticle transmission and Andreev reflection, where electron–hole conversion transfers charge through the superconducting condensate. Using the Anantram–Datta scattering formalism, we decompose the charge current and zero-frequency noise into Andreev, quasiparticle, and interference contributions. Although the interference term prevents a simple additive bound at the level of individual noise components, we show that the nonequilibrium excess noise admits a positive representation. This allows us to prove a hybrid quantum thermodynamic uncertainty relation valid for an arbitrary real superconducting gap. Our result extends the pure-Andreev quantum TUR to regimes where quasiparticle and Andreev processes coexist, clarifying how superconducting coherence reshapes current fluctuations while preserving a universal dissipation–precision constraint in hybrid quantum conductors.
\end{abstract}

\maketitle

%% -----------------------------------------------------------------------
\emph{Introduction.}---
The relationship between current fluctuations and dissipation is a
cornerstone of nonequilibrium statistical mechanics. In classical
Markovian systems far from equilibrium, this relationship is sharply
quantified by the \emph{Thermodynamic Uncertainty Relation}
(TUR)~\cite{Barato2015,Gingrich2016}: the relative fluctuations of
any steady-state current are bounded below by $2k_{\rm B}/\sigma$,
where $\sigma$ is the entropy production rate. The TUR therefore
encodes a fundamental trade-off between thermodynamic cost and the
precision with which a system can sustain a nonequilibrium current,
with broad implications for molecular motors, biochemical networks,
and autonomous thermal machines~\cite{Horowitz2020,Seifert2012}.
Extending the TUR to the quantum domain implies considering the wave nature of matter and non-Markovian dynamics in many scenarios. Under such perspective, quantum coherence can tighten or loosen the classical
precision--dissipation trade-off depending on the transmission spectrum~\cite{Brandner2018,Ptaszynski2019,Hasegawa2019}. Indeed, very recently a quantum formulation for the
TURs (QTUR) has been provided which is applicable for quantum conductors attached to metallic contacts \cite{Brandner2025}. Including hybrid Normal--Superconductor (N--S) junctions introduce an
additional layer of complexity providing new transport channels that the QTUR cannot properly describe. In N-S junctions, when a normal metallic contact is proximitized by a
superconductor, subgap transport (for energies below the superconducting gap denoted by $\Delta$) is dominated by \emph{Andreev
reflection}~\cite{Andreev1964,Blonder1982}. Here, an incident electron is
retroreflected as a hole, and a Cooper pair is injected into the
condensate. This process doubles the effective charge of the carriers,
profoundly altering both mean currents and their
fluctuations~\cite{Beenakker1992,Lambert1991}. Above the gap,
quasiparticle transmission and Andreev processes
coexist~\cite{Anantram1996,deFranceschi2010}, and the interplay
between the two channels governs the transport through the N-S junction. More intricate physics arises when a quantum dot (QD) is inserted between the electrodes enables
resonant Andreev tunneling, making the transmission probabilities
tunable via a gate voltage~\cite{FazioRaimondi1998,Governale2008NS}.
%% -----------------------------------------------------------------------
\begin{figure}[t]
  \centering
  \includegraphics[width=\linewidth]{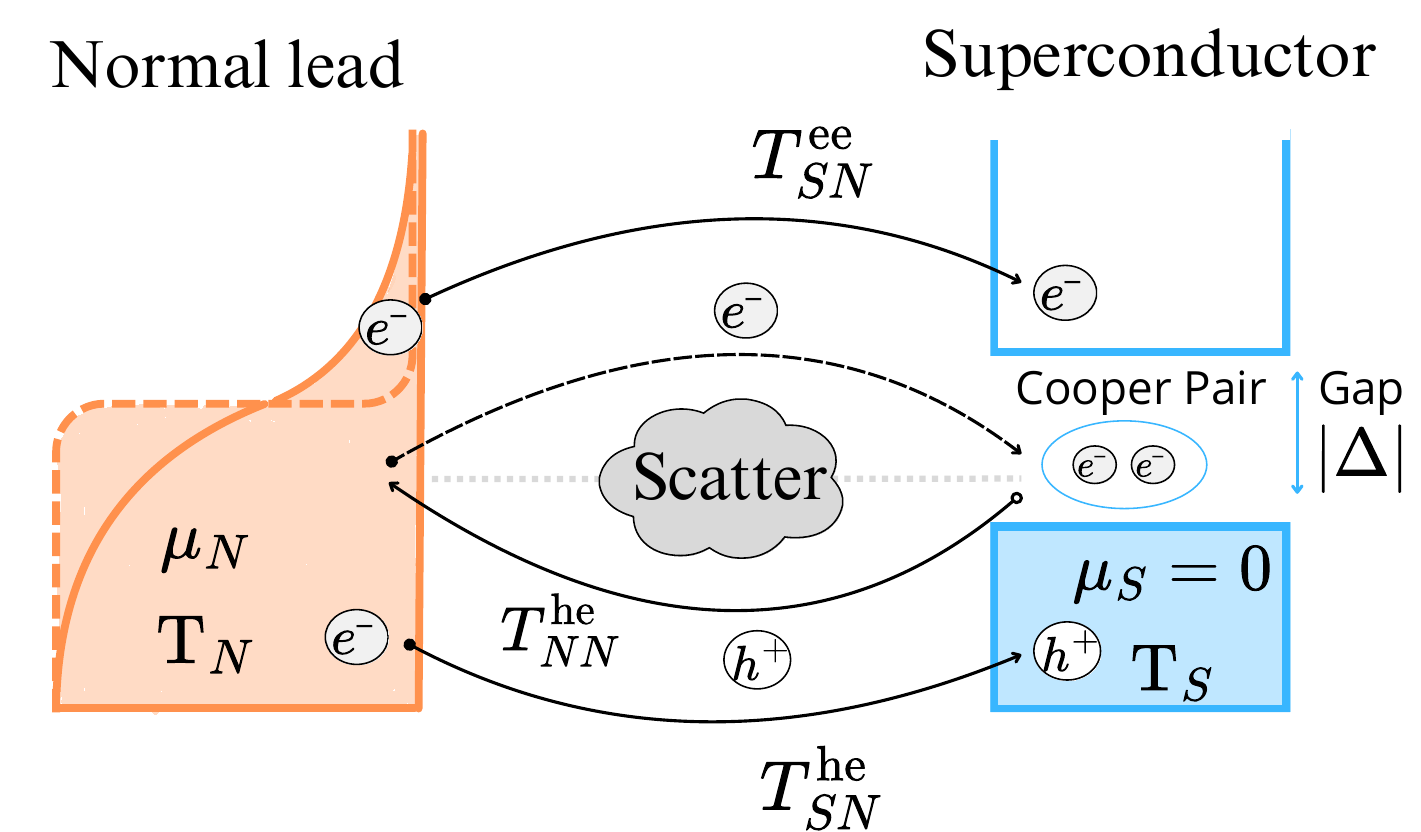}
  \caption{Schematic of the normal--superconductor hybrid device.  A
  scattering region is coupled to a normal reservoir (temperature
  $T_N$, chemical potential $\mu_N$) and a grounded superconductor
  (gap $\Delta$, temperature $T_S$, $\mu_S=0$).  Two competing
  transport channels coexist: (i) \emph{Andreev reflection} where an
  electron from the normal lead is converted into a retroreflected
  hole, transferring a Cooper pair into the condensate with probability
  $T_{NN}^{\rm he}$, and (ii) the \emph{quasiparticle transmission} in which an
  electron from the normal lead is transmitted to the superconductor
  either as a quasielectron (probability $T_{SN}^{\rm ee}$) or as a
  quasihole (probability $T_{SN}^{\rm he}$).}
  \label{fig:Sketch}
\end{figure}

Despite significant progress in characterizing transport and correlations in N–S hybrid devices, the formulation of a thermodynamic bound for these systems has not been rigorously addressed. Firstly, a large deviation from the classical TUR was reported 
for N--S and S--S junctions~\cite{Cuevas}. Such departure from the classical TUR was attributed to additional transport channels such as multiple Andreev reflections. Later on, and within 
a perturbative scheme (in the low-voltage regime) a quantum hybrid TUR (HQTUR) was derived, resulting in a renormalized QTUR \cite{Brandner2025} in which the electron charge $e$ is replaced by $Ne$, where $N$ is the 
highest charge quantum involved in N--S or S--S transport~\cite{Ohnmacht2025}. The validity of such HQTUR was  tested numerically for higher voltages and temperatures.  Besides, in the purely Andreev regime, for $\Delta \to \infty$, the same HQTUR was derived for N--S junctions~\cite{Governale2026,Tesser2022}. 
In short, there is  no rigorous derivation for the thermodynamic bound in N-S systems that consider arbitrary
bias, temperature, and finite superconducting gap.

Our description is based on the Anantram--Datta scattering
formalism where the full charge current noise is computed \cite{Anantram1996}. We show that $\mathcal{S}$ can be written as three separate contributions,
namely, the Andreev noise $\mathcal{S}_A$, the quasiparticle part denoted 
as $\mathcal{S}_Q$ and, an interference term 
$\mathcal{S}_{\rm Cross}$, which appears only under nonequilibrium conditions. 
The full noise can also be partitioned in thermal $\mathcal{S}^{\mathrm{th}}$ and shot noise $\mathcal{S}^{\mathrm{sh}}$ parts. 
From the positivity of the total shot noise we  provide a rigorous,
non-perturbative proof of the hybrid quantum TUR valid for
arbitrary superconducting gap as
\begin{equation}
  \sigma \;\geq\; \frac{k_{\rm B}|I|}{e}\,
  \operatorname{arsinh}\!\left(\frac{2e|I|}{\mathcal{S}}\right),
  \label{eq:HQTUR_main}
\end{equation}
where $I$ is the charge current.  The key technical contribution is an
exact algebraic identity showing that the total nonequilibrium shot
noise $\mathcal{S}_{\rm sh}$ is non-negative despite the fact that the interference term
$\mathcal{S}_{\rm Cross}$ attains negative values.  

\emph{Model and formalism.}---
We consider a coherent scatterer tunnel-coupled to a normal fermionic
reservoir and a superconducting electrode (Fig.~\ref{fig:Sketch}).
Working in the Nambu (electron--hole) basis, the scattering matrix
elements $s_{ij}^{\alpha\beta}$ give the amplitude for a particle of
type $\beta\in\{e,h\}$ incoming from lead $j\in\{N,S\}$ to leave as
type $\alpha\in\{e,h\}$ into lead $i$.  The corresponding
transmission probabilities are $T_{ij}^{\alpha\beta}(\epsilon)=
|s_{ij}^{\alpha\beta}|^2$.

Within the Anantram--Datta formalism~\cite{Anantram1996}, the charge
current at the normal terminal reads ($I_N=I$)
\begin{equation}
  I = \frac{e}{h}\int_{-\infty}^{\infty}\!\mathrm{d}\epsilon
  \sum_{\alpha\beta j}\mathrm{sgn}(\alpha)
  \bigl[\delta_{Nj}\delta_{\alpha\beta}-T_{Nj}^{\alpha\beta}(\epsilon)\bigr]
  f_{j\beta}(\epsilon),
  \label{eq:currentgeneral}
\end{equation}
where $\mathrm{sgn}(e)=+1$, $\mathrm{sgn}(h)=-1$, and
$f_{ie}(\epsilon)=f_i(\epsilon-\mu_i)$,
$f_{ih}(\epsilon)=f_i(\epsilon+\mu_i)$ are the electron and hole
Fermi--Dirac distributions of lead $i$ at temperature $T_i$.  The
superconductor is grounded, $\mu_S=0$.

We identify the \emph{Andreev
transmission probability} as $T_A(\epsilon)=T_{NN}^{he}(\epsilon)=T_{NN}^{eh}$
that satisfies $T_A(\epsilon)=T_A(-\epsilon)$. The \emph{quasiparticle
transmission} $T_Q(\epsilon)=T_{SN}^{ee}(\epsilon)+T_{SN}^{he}(\epsilon)$
accounts for electrons and holes transmitted into the superconductor
as quasiparticles.  Flux conservation imposes $T_{NN}^{ee}(\epsilon)=1-T_A(\epsilon)-T_Q(\epsilon)$. Expressing the current in terms of $T_A$ and $T_Q$ yields the
physically transparent form
\begin{equation}
  I = \frac{2e}{h}\int_{-\infty}^{\infty}\!\mathrm{d}\epsilon
  \bigl[T_A(f_{Ne}-f_{Nh})+T_Q(f_{Ne}-f_S)\bigr],
  \label{eq:current_TATQ}
\end{equation}
An analogous expression
holds for the energy current $J^E$
\begin{equation}
  J^E = \frac{2}{h}\int_{-\infty}^{\infty}\!\mathrm{d}\epsilon
 \,\epsilon \,\bigl[T_A(f_{Ne}-f_{Nh})+T_Q(f_{Ne}-f_S)\bigr],
  \label{eq:energy_current}
\end{equation}
Crucially, the Andreev channel carries
no energy due to particle-hole symmetry.  The entropy production rate follows Clausius relation,
\begin{equation}
  \sigma = -\frac{J_N}{T_N}-\frac{J_S}{T_S},
  \label{eq:sigma}
\end{equation}
where $J_i=J_i^E-(\mu_i/e)I_i$, with $i\in \{N,S\}$ are the heat currents.

%% -----------------------------------------------------------------------
\emph{Decomposition of the current noise.}---
The zero-frequency charge noise $\mathcal{S}$ for the normal charge current
is computed from the Anantram--Datta
formula~\cite{Anantram1996} as
$\mathcal{S}=\mathcal{S}^{AA}+\mathcal{S}^{AB}$, with
\begin{widetext}
\begin{align}
\mathcal{S}^{AA}
&=\frac{e^2}{h}\int_{-\infty}^{\infty}\!\mathrm{d}\epsilon\,
\sum_{\alpha}\Bigl\{
  \bigl[1-T_{NN}^{\alpha\alpha}(\epsilon)\bigr]^2
  f_{N\alpha}(1-f_{N\alpha})
  +\!\!\sum_{(k\gamma l\delta)\neq(N\alpha N\alpha)}\!\!
  T_{Nk}^{\alpha\gamma}T_{Nl}^{\alpha\delta}
  f_{k\gamma}(1-f_{l\delta})
\Bigr\},
\label{eq:SAA}\\
\mathcal{S}^{AB}
&=\frac{e^2}{h}\int_{-\infty}^{\infty}\!\mathrm{d}\epsilon\,
\sum_{\alpha}\Bigl\{
  2T_{NN}^{\alpha\bar\alpha}f_{N\bar\alpha}(1-f_{N\bar\alpha})
  +\sum_{k\gamma}s_{Nk}^{\bar\alpha\gamma}s_{Nk}^{\alpha\gamma\dagger}
   f_{k\gamma}
   \sum_{l\delta}s_{Nl}^{\alpha\delta}s_{Nl}^{\bar\alpha\delta\dagger}
   f_{l\delta}
\Bigr\}.
\label{eq:SAB}
\end{align}
\end{widetext}
Expressing the full noise, in terms of the Andreev and quasiparticle transmissions, i.e., $T_A$ and $T_Q$ reveals three physically
distinct contributions, 
\begin{equation}
\mathcal{S}=\mathcal{S}_A+\mathcal{S}_Q
+\mathcal{S}_{\rm Cross} 
\end{equation}
 whose separate roles are central to the
proof of the HQTUR.  The \emph{Andreev noise}
\begin{align}
\mathcal{S}_A &= \frac{4e^2}{h}\int_{-\infty}^{\infty}\!\mathrm{d}\epsilon\,
\Bigl\{T_A\bigl[f_{Ne}(1-f_{Ne})+f_{Nh}(1-f_{Nh})\bigr]\notag\\
&\qquad\qquad +T_A(1-T_A)(f_{Ne}-f_{Nh})^2\Bigr\}
\label{eq:SA}
\end{align}
is manifestly non-negative ($\mathcal{S}_A\geq 0$) and has the same
structure as the noise of a normal conductor with doubled charge.
The factor of 4 (instead of 2) reflects the $2e$ charge quanta
transferred by each Andreev event.  The \emph{quasiparticle noise}
\begin{align}
\mathcal{S}_Q &= \frac{2e^2}{h}\int_{-\infty}^{\infty}\!\mathrm{d}\epsilon\,
\Bigl\{T_Q\bigl[f_{Ne}(1-f_{Ne})+f_S(1-f_S)\bigr]\notag\\
&\qquad\qquad +T_Q(1-T_Q)(f_{Ne}-f_S)^2\Bigr\}
\label{eq:SQ}
\end{align}
is also non-negative ($\mathcal{S}_Q\geq 0$) and recovers the
standard noise formula in the normal limit $\Delta\to 0$.  The
\emph{cross (interference) term}
\begin{widetext}
\begin{multline}
\mathcal{S}_{\rm Cross}
=\frac{2e^2}{h}\int_{-\infty}^{\infty}\!\mathrm{d}\epsilon\,
\Bigl\{
-T_AT_Q\bigl[(f_{Ne}-f_{Nh})^2+2(f_{Ne}-f_{Nh})(f_{Ne}-f_S)\bigr]\\
+\bigl|s_{NN}^{he}s_{NN}^{ee\dagger}+s_{NN}^{hh}s_{NN}^{eh\dagger}\bigr|^2
 (f_{Nh}-f_S)^2
+2\operatorname{Re}\!\bigl[s_{NN}^{he}s_{NN}^{ee\dagger}
\bigl(s_{NN}^{he}s_{NN}^{ee\dagger}+s_{NN}^{hh}s_{NN}^{eh\dagger}\bigr)^\dagger\bigr]
(f_{Ne}-f_{Nh})(f_{Nh}-f_S)
\Bigr\}
\label{eq:SCross}
\end{multline}
\end{widetext}
arises from quantum interference between the Andreev and
quasiparticle amplitudes. Unlike $\mathcal{S}_{A,Q}$, it attains negative values, and it
vanishes identically in both limits $\Delta\to 0$ (no Andreev
reflection) and $\Delta\to\infty$ (no quasiparticle transmission).

The current and noise expressions derived here within scattering theory admit an equivalent formulation in terms of nonequilibrium Green functions, with the correspondence between the two approaches provided by the Fisher--Lee relation~\cite{Fisher1981}. The equivalence is established in Appendix~\ref{app:green-functions}, in the section entitled ``Equivalence between the scattering-matrix and Green's-function formalisms.''

\begin{figure}[t]
  \centering
  \includegraphics[width=\linewidth]{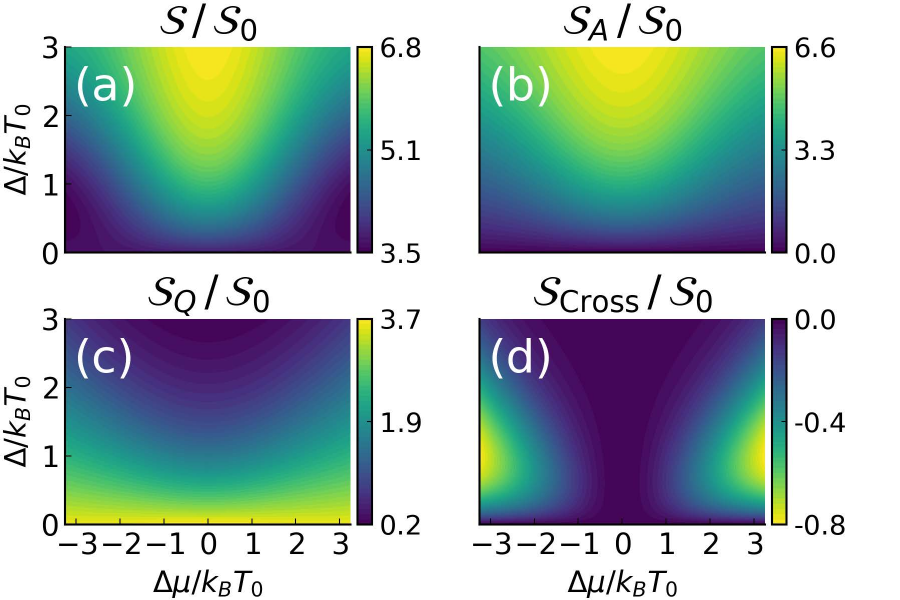}
  \caption{Color maps of the zero-frequency charge noise versus
  applied bias $\Delta\mu/k_BT_0$ and superconducting gap
  $\Delta/k_BT_0$ for a normal--superconductor quantum-dot junction.
  (a) Total noise $\mathcal{S}$.  (b) Andreev part $\mathcal{S}_A$.
  (c) Quasiparticle part $\mathcal{S}_Q$.  (d) Interference term
  $\mathcal{S}_{\rm Cross}$.  In the normal limit $\Delta\to 0$ only
  $\mathcal{S}_Q$ survives; as $\Delta$ grows, $\mathcal{S}_A$
  dominates in the subgap regime, while $\mathcal{S}_{\rm Cross}$
  becomes finite in the crossover region where both channels are
  active.  Parameters: $\mathcal{S}_0 = e^2/h$, $k_BT_{N,S}=k_BT_0 = 1$, $\epsilon_d=0$,
  $h\Gamma_{N,S}=6\,k_BT_0$.}
  \label{fig:noise_colormap}
\end{figure}

Charge current fluctuations are  illustrated in Fig. \ref{fig:noise_colormap}. Here, the full current noise $\mathcal{S}$ versus the bias voltage $\Delta\mu/k_BT_0$ and superconducting gap $\Delta/k_BT_0$ (with $T_0$ as the temperature for both contacts) is plotted in Fig. \ref{fig:noise_colormap}(a). The three contributions to the full noise, $\mathcal{S}_A$, $\mathcal{S}_Q$, $\mathcal{S}_\mathrm{Cross}$   are  shown in Fig. \ref{fig:noise_colormap}(b)-(d). The Andreev part $\mathcal{S}_A$  is displayed in Fig. \ref{fig:noise_colormap}(b) showing higher values for large $\Delta$. The quasiparticle
contribution $S_Q$ shown in Fig. \ref{fig:noise_colormap}(c)  runs in opposite direction, it is larger for small $\Delta$. Finally,  the cross (interference) term $\mathcal{S}_\mathrm{Cross}$ depicted in Fig. \ref{fig:noise_colormap}(d) displays negative values, with large absolute values far from equilibrium (large $\Delta\mu/k_BT_0$) and moderate $\Delta$. Notice, that the
normal limit considered as $\Delta\to 0$ shows that the Andreev reflection is fully suppressed
and only $S_Q$ survives, recovering the standard quantum conductor
result.

\emph{Hybrid Quantum Thermodynamic Uncertainty Relation---} \label{positivity shot noise}
\begin{figure}[t]
    \centering
    \includegraphics[width=\linewidth]{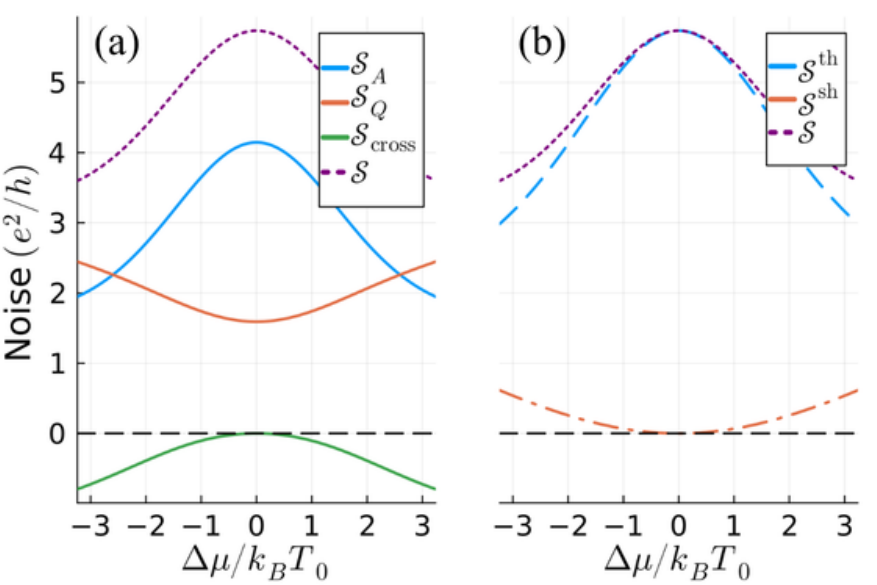}
    \caption{(a) Noise versus the applied bias voltage. Zero-frequency noise for $\mathcal{S}$, and the individual contributions, namely, the Andreev part $\mathcal{S}_{A}$, the quasiparticle  noise $\mathcal{S}_Q$ and the cross term $\mathcal{S}_{\mathrm{Cross}}$, which displays negative values. (b) Illustration of the total noise, its thermal and shot noise parts as indicated. Parameters: same as Fig. \ref{fig:noise_colormap}.}
     \label{fig:S th and S sh}
\end{figure}
The demonstration of the HQTUR rests on the positivity of the total
charge noise. To establish this, it is convenient to decompose
$\mathcal{S}$ into its thermal and shot noise contributions,
$\mathcal{S} = \mathcal{S}^{\mathrm{th}} + \mathcal{S}^{\mathrm{sh}}$,
where $\mathcal{S}^{\mathrm{sh}}$ vanishes at equilibrium and becomes
finite only under nonequilibrium driving.
Figure~\ref{fig:S th and S sh}(a) shows the three physically distinct
contributions to the total noise, namely, the Andreev part
$\mathcal{S}_A \geq 0$, the quasiparticle contribution
$\mathcal{S}_Q \geq 0$, and the cross (interference) term
$\mathcal{S}_{\mathrm{Cross}}$.
Although the not positive-semidefinite character of $\mathcal{S}_{\mathrm{Cross}}$ prevents a
simple additive bound at the level of individual components,
reorganizing the noise into thermal and shot parts reveals that both
are manifestly non-negative, $\mathcal{S}^{\mathrm{th}} \geq 0$ and
$\mathcal{S}^{\mathrm{sh}} \geq 0$, as shown in
Fig.~\ref{fig:S th and S sh}(b). This structure follows from the
fact that the thermal noise receives contributions solely from the
thermally activated Andreev and quasiparticle channels,
$\mathcal{S}_{A,Q} = \mathcal{S}_{A,Q}^{\mathrm{th}} +
\mathcal{S}_{A,Q}^{\mathrm{sh}}$, while the cross term is purely
a nonequilibrium effect,
$\mathcal{S}_{\mathrm{Cross}} = \mathcal{S}_{\mathrm{Cross}}^{\mathrm{sh}}$
with $\mathcal{S}_{\mathrm{Cross}}^{\mathrm{th}} = 0$.

For completeness, we write below their explicit expressions to make clearer the previous discussion
\begin{equation}
\begin{gathered}
 \mathcal{S}_{A,Q}^{\mathrm{th}}   = g_{A,Q}\frac{e^{2}}{h} \int_{-\infty}^{\infty}\mathrm{d}\epsilon \, T_{A, Q} \big[ f_{N\text{e}}(1-f_{N\text{e}}) \\ + f_{N\text{h}, S}(1-f_{N\text{h}, S}) \big]
\end{gathered}
\end{equation}
and
\begin{equation}
    \mathcal{S}_{A,Q}^{\mathrm{sh}} = g_{A,Q}\frac{e^{2}}{h} \int_{-\infty}^{\infty}\mathrm{d}\epsilon \, T_{A, Q} (1-T_{A,Q}) \left( f_{N\text{e}}- f_{N\text{h}, S} \right)^2
\end{equation}
with $g_A = 4$ and $g_Q = 2$. Notice that $\mathcal{S}_{A,Q}^{\mathrm{sh}} \geq 0$.  After some algebra, and denoting $x = f_{N\text{e}}- f_{N\text{h}} $, $y = f_{N\text{e}}- f_{ S}$, $z = y - x = f_{N\text{h}}- f_{ S}$, $a = s_{NN}^{\text{he}}s_{NN}^{\text{ee}\dagger}$, $b = s_{NN}^{\text{hh}}s_{NN}^{\text{eh}\dagger}$, and $ c = a + b$, then we can write the total shot noise, i.e., $\mathcal{S}^{\mathrm{sh}}=\mathcal{S}_{A}^{\mathrm{sh}}+\mathcal{S}_{Q}^{\mathrm{sh}}+\mathcal{S}_{\mathrm{ Cross}}$
\begin{widetext}
\begin{gather}
\begin{gathered}
\mathcal{S}^{{\mathrm{sh}}}
=
\frac{2e^{2}}{h}
\int_{-\infty}^{\infty}\mathrm{d}\epsilon
\left\{
\left|cz+ax\right|^{2}
+
T_{Q}\left(1-T_{Q}\right)
\left(
y-\frac{T_{A}}{1-T_{Q}}x
\right)^{2}
\right.
\left.
+
\frac{T_{A}\left(1-T_{A}-T_{Q}\right)}
{1-T_{Q}}
x^{2}
\right\}.
\end{gathered}
\end{gather}
\end{widetext}
Since $0 \leq T_{A,Q} \leq 1$ and $0 \leq |s_{NN}^{\text{ee}}|^2 = 1 - T_A - T_Q \leq 1$, then $\mathcal{S}^{sh} \geq 0$. (In the limiting case $T_Q \to 1$ the second and third terms vanish while the first remains non-negative, so positivity holds throughout.) This way, in hybrid Normal--Superconducting devices, the relation $\mathcal{S} = \mathcal{S}^{\rm{th}} + \mathcal{S}^{\rm{sh}}   \geq \mathcal{S}^{\rm{th}}  = \mathcal{S}^{\rm{th}} _A + \mathcal{S}^{\rm{th}} _Q$ is also satisfied. 

Notice that we also have that $\mathcal{S}_{A,Q}^{\rm{th}} \geq 0$ as $0 \leq f_i \leq 1$, so $\mathcal{S}^{\rm th} \geq 0$. 
\begin{figure}
    \centering
    \includegraphics[width=\linewidth]{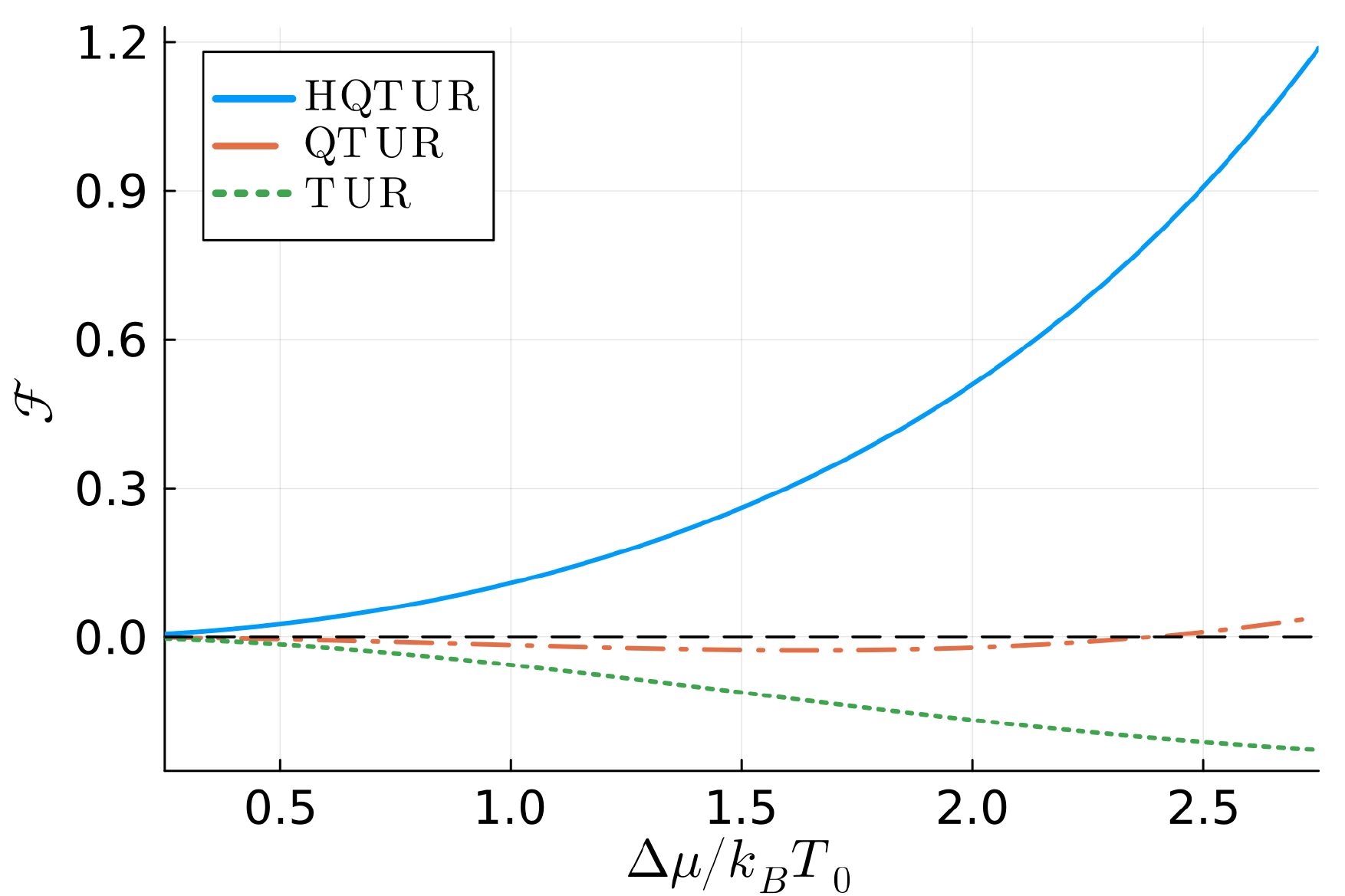}
    \caption{Thermodynamic uncertainty relations, namely, TUR, QTUR, and HQTUR as a function of the electric bias applied to the hybrid normal--superconductor quantum dot system. The dimensionless function $\mathcal{F}$ characterizes the bound that every thermodynamical uncertainty relation establishes, such that $\mathcal{F} < 0$ means that the relation is violated. Only the HQTUR is not violated for the full range of applied bias voltages. Parameters: Same as Fig. \ref{fig:noise_colormap}.}
    \label{fig:HQTUR}
\end{figure}
Following the derivation for the QTUR, but now generalized to the hybrid NS case, we establish that
\begin{equation}
\begin{gathered}
        \sigma_Q \geq \frac{2k_B \lvert I_Q \rvert}{e} \, \operatorname{arsinh}\!\left( \frac{e \lvert I_Q \rvert}{\mathcal{S}^{\text{th}}_{Q}} \right) \\
       \geq \frac{k_B \lvert I_Q \rvert}{e} \, \operatorname{arsinh}\!\left( \frac{2e \lvert I_Q \rvert}{\mathcal{S}^{\text{th}}_{Q}}\right).\nonumber
    \label{eq:HQTUR Q}
\end{gathered}
\end{equation}
where in the last step we used that $2 \operatorname{arsinh}(x) \geq \operatorname{arsinh}(2x)$ for all $x \geq 0$. Additionally, the Andreev contribution fulfills, \cite{Governale2026}
\begin{equation}
    \sigma_A \geq \frac{k_B \lvert I_A \rvert}{e} \, \operatorname{arsinh}\!\left( \frac{2e \lvert I_A \rvert}{\mathcal{S}^{\text{th}}_{A}} \right) .
    \label{HQTUR A}
\end{equation}

Adding both inequalities, with $\sigma =\sigma_A+\sigma_Q$

\begin{equation}
\begin{gathered}
         \sigma \geq \frac{k_B \lvert I_Q \rvert}{e} \, \operatorname{arsinh}\!\left( \frac{2e \lvert I_Q \rvert}{\mathcal{S}^{\text{th}}_{Q}}\right) \\
        + \frac{k_B \lvert I_A \rvert}{e} \, \operatorname{arsinh}\!\left( \frac{2e \lvert I_A \rvert}{\mathcal{S}^{\text{th}}_{A}} \right) .
    \label{eq:HQTUR sum}
\end{gathered}
\end{equation}

Using that $|I_A| + |I_Q| \geq |I_A + I_Q|$, $\operatorname{arsinh}(x) \geq \operatorname{arsinh}(y)$ for all $x \geq y \geq 0$, the convexity of $\phi(x) = x  \operatorname{arsinh}\!\left( 2e x \right)$ and $\mathcal{S}^{\text{th}}_{A} + \mathcal{S}^{\text{th}}_{Q} \geq \mathcal{S}^{\text{th}}_{A, Q} \geq 0$ as $\mathcal{S}^{\text{th}}_{A,Q} \geq 0$, so $1/\mathcal{S}^{\text{th}}_{A, Q} \geq 1/(\mathcal{S}^{\text{th}}_{A} + \mathcal{S}^{\text{th}}_{Q}) \geq 0$, then
\begin{equation}
\begin{gathered}
    \sigma \geq \frac{k_B \lvert I_A + I_Q \rvert}{e} \, \operatorname{arsinh}\!\left( \frac{2e \lvert I_A + I_Q \rvert}{\mathcal{S}^{\text{th}}_{A} + \mathcal{S}^{\text{th}}_{Q}} \right)\\
    = \frac{k_B \lvert I\rvert}{e} \, \operatorname{arsinh}\!\left( \frac{2e \lvert I \rvert}{\mathcal{S}^{\text{th}}} \right).
\end{gathered}
    \label{HQTUR Sth}
\end{equation}
Employing $\mathcal{S} \geq \mathcal{S}^{\text{th}} \geq 0$, so $1/\mathcal{S}^{\text{th}} \geq 1/\mathcal{S} \geq 0$ and as $\operatorname{arsinh}(x) \geq \operatorname{arsinh}(y)$ for all $x \geq y \geq 0$, the HQTUR is rigorously demonstrated, and reads as
\begin{equation}
    \sigma \geq \frac{k_B \lvert I\rvert}{e} \, \operatorname{arsinh}\!\left( \frac{2e \lvert I \rvert}{\mathcal{S}} \right).
    \label{HQTUR}
\end{equation}
This is the central result of our research, the Hybrid Quantum TUR valid for hybrid systems. Below we illustrate this new thermodynamic bound and compare with the classical bound, the TUR, and its generalization for normal quantum systems, the QTUR. 

\emph{Validity of the TUR, QTUR and HQTUR---} We numerically tested in Fig. \ref{fig:HQTUR}. These three dimensionless functions are
\begin{align}
  \mathcal{F}_{\mathrm{TUR}}   & = \frac{S\,\sigma}{2k_B I^2} - 1,
  \label{eq:FTUR}\\
  \mathcal{F}_{\mathrm{QTUR}}  & = \frac{S}{e|I|}
      \sinh\!\left(\frac{e\sigma}{2k_B|I|}\right) - 1,
  \label{eq:FQTUR}\\
  \mathcal{F}_{\mathrm{HQTUR}} & = \frac{S}{2e|I|}
      \sinh\!\left(\frac{e\sigma}{k_B|I|}\right) - 1,
  \label{eq:FHQTUR}
\end{align}
A violation of any of this bound is indicated as $\mathcal{F}_X<0$. 
More precisely, Fig.~\ref{fig:HQTUR} shows that both the classical TUR
and the QTUR are violated over broad voltage windows, where Andreev and quasiparticle channels interfere
most strongly. The HQTUR, by contrast, satisfies
$\mathcal{F}_\mathrm{HQTUR}\geq 0$ across the entire range of
parameters explored, confirming the validity at arbitrary superconducting gap $\Delta$. Together, these results
establish the HQTUR as the natural and tight
dissipation--precision constraint for hybrid
normal--superconducting quantum conductors, extending the
previously known pure-Andreev limit to the full transport
regime where quasiparticle processes and Cooper-pair correlations
coexist.

%% -----------------------------------------------------------------------
\emph{Conclusions.}---
We have derived and rigorously proved the Hybrid Quantum Thermodynamic
Uncertainty Relation.  The result extends the
pure-Andreev quantum TUR~\cite{Governale2026} and the normal-conductor
QTUR~\cite{Brandner2018} to the full N--S transport regime, where
Andreev reflection and quasiparticle transmission coexist.  The proof
rests on an exact decomposition of the zero-frequency noise into
Andreev, quasiparticle, and interference parts, together with an
algebraic identity showing that the total nonequilibrium shot noise is
non-negative despite the not positive-semidefinite interference contribution $\mathcal{S}_{\rm Cross}$. Besides, we have illustrated the fulfillment of the HQTUR for NS systems by numerical simulations for which the classical TUR and the quantum TUR fail.
Our result establishes the HQTUR as the natural and tight
dissipation--precision constraint for hybrid normal--superconducting
quantum conductors.  

\acknowledgments{
We thank Alba Mayor--Fernández, Gianmichele Blasi and Sungguen Ryu for fruitful discussions.
We acknowledge support from the Spanish State Research Agency
(MCIN/AEI/10.13039/501100011033) and FEDER (EU) under
Grant No.~PID2020-117347GB-I00, and the Mar\'ia de Maeztu Project
No.~CEX2021-001164-M.
}

\onecolumn\newpage
\appendix
\section*{Appendix}
\setcounter{equation}{0}
\setcounter{figure}{0}
\setcounter{table}{0}
\renewcommand{\theequation}{A\arabic{equation}}
\renewcommand{\thefigure}{A\arabic{figure}}
\renewcommand{\thetable}{A\arabic{table}}

\section{Physical framework}
\label{app:physical-framework}

\label{sec:framework}

In this section we obtain the expressions for the charge current and charge noise, as well as the energy and heat currents, so all these quantities can be related in the hybrid quantum thermodynamic uncertainty relation (HQTUR)~\cite{Barato2015,Gingrich2016,Horowitz2020,Brandner2025,Governale2026}.

We consider a central structure, modeled as a scatterer, coupled to two leads consisting of macroscopic reservoirs of fermionic particles in thermodynamic equilibrium. One lead is normal, i.e., a fermionic ideal gas, with occupation levels given by the Fermi--Dirac distribution in Eq.~\eqref{eq:fermi}, with temperature $\mathrm{T}_N$ and chemical potential $\mu_N$; the other lead is superconducting, with order parameter $\Delta$, temperature $\mathrm{T}_S$ and chemical potential $\mu_S = 0$. Throughout this work, all thermodynamic parameters and the bias between the two leads are taken to be constant (time-independent), and the system is described in the non-equilibrium steady state (NESS), characterized by stationary particle, charge, and energy currents, and reached after transient effects have decayed.

We assume absence of additional interactions, namely zero external magnetic field and lack of electron--electron interactions beyond the effective mean-field BCS pairing description. Furthermore, the mean-field description, given by equation~\eqref{Superconducting Hamiltonian equation}, is invariant under a global $U(1)$ transformation, under which the fermionic operators transform as $c_{\alpha k\sigma}\mapsto e^{i\chi}c_{\alpha k\sigma}$, with $\chi \in \mathbb{R}$, while the superconducting order parameter transforms as $\Delta \mapsto e^{2i\chi}\Delta$, given that $\Delta$ is defined in the mean-field BCS theory as $\Delta := \sum_{k} g \, \langle c_{S,-k\downarrow} \, c_{S,k\uparrow} \rangle$, with $g >0$ and $c_{S, \pm k \sigma}$ fermionic annihilation operators. Since the setup contains a single stationary superconducting reservoir in thermodynamic equilibrium, writing $\Delta=|\Delta|e^{i\phi}$, its global condensate phase can be removed by a global $U(1)$ transformation, selecting the correct phase $\chi$ of this $U(1)$ transformation, i.e., $2\chi + \phi = n \pi$ with $n \in \mathbb{N} \cup \{ 0\}$. We therefore choose a real superconducting order parameter,
\begin{equation}
\Delta^*=\Delta,
\end{equation}
whose magnitude $|\Delta|$ is an arbitrary model parameter. Hence, $\Delta$ can take any real value, $\Delta = \pm |\Delta|$. For brevity, we refer to $\Delta$ as the superconducting gap throughout the remainder of this work, since $\Delta = \pm |\Delta|$, so $\Delta$ essentially coincides with the superconducting gap magnitude \(\lvert\Delta\rvert\) up to an overall sign. \\

In addition, the number of particles remain invariant under this $U(1)$ global symmetry, since the particle number operator in $\alpha$ is given by $N_\alpha = \sum_{k \sigma} c_{\alpha k\sigma}^\dagger c_{\alpha k\sigma}$, and transforms as $c_{\alpha k\sigma}^\dagger c_{\alpha k\sigma} \mapsto e^{-i\chi} c_{\alpha k\sigma}^\dagger e^{i\chi} c_{\alpha k\sigma} = c_{\alpha k\sigma}^\dagger c_{\alpha k\sigma}$. Thus, the charge current operator $I_\alpha = -e \frac{d N_\alpha}{dt}$ remains invariant too. Here, $e$ is the elementary charge, and $I_\alpha$ is positive if the particles flow out from $\alpha$. Since the Hamiltonian is invariant under this transformation, the energy current is invariant, so the heat current is also unaffected by $U(1)$ transformations.

Finally, we assume one transport channel for each lead. Each channel can carry both electrons and holes, inducing an electron channel and a hole channel in both leads.

\subsection{Scattering matrix}
\label{sec:scattering_matrix}

In a general framework, the scattering operator~\cite{Buttiker1992,Datta1995,Blanter2000} relates all the initial states $\{\phi\}$ with a final state $\psi$, for each final state, such that
\begin{equation}
    \psi = \sum_{\phi} s_{_{\psi \phi}}\, \phi,
    \label{eq:scattering}
\end{equation}
where the operators $s_{_{\psi\phi}}$ are called scattering-operator elements and are linear operators that represent the amplitude for an initial state $\phi$ to leave the scatterer as a final state $\psi$. Additionally, $s_{_{\psi\phi}}\phi$ denotes, in general, $s_{_{\psi\phi}}$ acting on $\phi$ as an operator, rather than an ordinary multiplication. Expressing the states in a coordinate representation in a basis, the scattering-operator elements can be represented by complex scalars, so the full scattering operator can be written as a complex matrix, called the scattering matrix. Equation~\eqref{eq:scattering} adopts the matrix form
\begin{equation}
    \hat{\psi}=\bm{s}\hat{\phi},
\end{equation}
where $\bm{s}$ is the scattering matrix and $\hat\phi,\hat\psi$ are the vectors of initial and final states, respectively.

We work in the Nambu basis~\cite{Nambu1960,PhysRevB.53.16390,Beenakker1997}
\begin{equation}
   \mathcal{B} :=  \{N\text{e},\,N\text{h},\,S\text{e},\,S\text{h}\},
\end{equation}
where each lead $\{N,S\}$ carries electron ($\text{e}$) and hole ($\text{h}$) channels. For a two-terminal N--S structure, the scattering matrix is arranged as
\begin{equation}
\bm{s}(\epsilon)=
\begin{pmatrix}
 s^{\text{ee}}_{{NN}} & s^{\text{eh}}_{NN} & s^{\text{ee}}_{NS} & s^{\text{eh}}_{NS} \\
 s^{\text{he}}_{{NN}} & s^{\text{hh}}_{{NN}} & s^{\text{he}}_{NS} & s^{\text{hh}}_{NS} \\
 s^{\text{ee}}_{{SN}} & s^{\text{eh}}_{{SN}} & s^{\text{ee}}_{SS} & s^{\text{eh}}_{SS} \\
 s^{\text{he}}_{{SN}} & s^{\text{hh}}_{{SN}} & s^{\text{he}}_{SS} & s^{\text{hh}}_{SS}
\end{pmatrix}.
\label{eq:scattering_matrix}
\end{equation}
Here $s_{ij}^{\alpha\beta}$ is the amplitude for an initial particle of type $\beta$ in lead $j$ to leave the scatterer as a final particle of type $\alpha$ in lead $i$. We denote the corresponding probabilities by
\begin{equation}
    T_{ij}^{\alpha\beta}(\epsilon)
    :=\left|s_{ij}^{\alpha\beta}(\epsilon)\right|^2.
    \label{eq:transmission}
\end{equation}
Probability conservation imposes unitarity~\cite{Buttiker1992,Datta1995,Beenakker1997,PhysRevB.53.16390},
\begin{equation}
\begin{gathered}
\bm{s}^{\dagger}(\epsilon)\bm{s}(\epsilon)
=\bm{s}(\epsilon)\bm{s}^{\dagger}(\epsilon)
=\mathbb{I},
\\
\sum_{\gamma,k}s_{ik}^{\alpha\gamma}(\epsilon)\,
\left[s_{jk}^{\beta\gamma}(\epsilon)\right]^{\dagger}
=\delta_{\alpha\beta}\delta_{ij},
\end{gathered}
\label{eq:unitarity}
\end{equation}
where $\mathbb{I}$ is the identity matrix. This way, taking $\alpha = \beta$ and $i = j$, then $0 \leq   \left|s_{ij}^{\alpha\beta}(\epsilon)\right|^2 \leq 1$, i.e, $0 \leq     T_{ij}^{\alpha\beta}(\epsilon) \leq 1$. It is useful to write explicitly the following identities from Eq.~\eqref{eq:unitarity}:
\begin{equation}
    T_{NN}^{\text{ee}}(\epsilon)
    +T_{NN}^{\text{eh}}(\epsilon)
    +T_{NS}^{\text{ee}}(\epsilon)
    +T_{NS}^{\text{eh}}(\epsilon)
    =1,
    \label{eq:unitarity_transmissions_explicit}
\end{equation}
\begin{equation}
    s_{NN}^{\text{he}}s_{NN}^{\text{ee}\dagger}
    +s_{NN}^{\text{hh}}s_{NN}^{\text{eh}\dagger}
    +s_{NS}^{\text{he}}s_{NS}^{\text{ee}\dagger}
    +s_{NS}^{\text{hh}}s_{NS}^{\text{eh}\dagger}
    =0.
    \label{eq:unitarity_scattering_explicit}
\end{equation}

Because we assume one single channel for each lead $i,j\in\{N,S\}$ and particle $\alpha,\beta\in\{\text{e},\text{h}\}$, then
\begin{equation}
    \left[s_{ij}^{\alpha\beta}(\epsilon)\right]^{\dagger}
    =\left[s_{ij}^{\alpha\beta}(\epsilon)\right]^{*}.
\end{equation}

For the conductors considered here, time-reversal symmetry implies the
microreversibility relation~\cite{Buttiker1992,Beenakker1997}, so the scattering matrix becomes symmetric:
\begin{equation}
    \bm{s}(\epsilon)=\bm{s}^{T}(\epsilon)
    \implies
    s_{ij}^{\alpha\beta}(\epsilon)
    =s_{ji}^{\beta\alpha}(\epsilon).
    \label{eq:symmetric_scattering_matrix}
\end{equation}
Thus, the diagonal elements in lead space ($i = j$) satisfy
\begin{equation}
    s_{ii}^{\alpha\beta}(\epsilon)
    =s_{ii}^{\beta\alpha}(\epsilon).
    \label{eq: microreversivility same contact}
\end{equation}

In addition, particle-hole symmetry is intrinsic to the
Bogoliubov-de Gennes description. In the Nambu basis, it imposes the
constraint~\cite{BlonderTinkhamKlapwijk1982,Beenakker1997,Fukui_2010}
\begin{equation}
    \bm{s}(\epsilon) = \mathcal{C} \, \bm{s}(-\epsilon)\, \mathcal{C}^{-1} = \tau_x \bm{s}^{*}(-\epsilon) \tau_x \implies s_{ij}^{\alpha\beta}(\epsilon)
    =\left[s_{ij}^{\bar\alpha\bar\beta}(-\epsilon)\right]^{*},
    \label{eq:PHS}
\end{equation}
where $\mathcal{C} = \tau_x K$ acts by performing complex conjugation (with $K$) and exchanging electrons and holes (with $\tau_x$), so
\begin{equation}
\begin{cases}
\bar\alpha,\bar\beta=\text{h} & \text{if } \alpha,\beta=\text{e},\\
\bar\alpha,\bar\beta=\text{e} & \text{if } \alpha,\beta=\text{h}.
\end{cases}
\end{equation}

\subsection{Charge current}
\label{sec:charge_current}

The charge current in hybrid conductors is given by the Anantram--Datta scattering formula~\cite{Anantram1996},
\begin{equation}
    I_N = \langle \hat{I}_N \rangle 
    =\frac{e}{h}\int_{-\infty}^{\infty}\mathrm{d}\epsilon
    \sum_{\alpha\beta j}\mathrm{sgn}(\alpha)
    \left[\delta_{Nj}\delta_{\alpha\beta}
    -T_{Nj}^{\alpha\beta}(\epsilon)\right]
    f_{j\beta}(\epsilon),
    \label{eq:charge_current}
\end{equation}
where $j\in\{N,S\}$ is a lead index and $\alpha,\beta\in\{\text{e},\text{h}\}$ are electron/hole indexes. The operator $\hat{I}_N$ denotes the charge current operator from Anantram--Datta scattering. We choose the sign convention where electrons flowing out of the contact contribute positively to the current, equivalently holes flowing into the contact. Hence
\begin{equation}
    \mathrm{sgn}(\text{e})=+1,
    \qquad
    \mathrm{sgn}(\text{h})=-1.
\end{equation}
The terms in Eq.~\eqref{eq:charge_current} represent electrons and holes flowing out of the contact. Equivalently, electrons flowing out of the lead contribute with charge $+e$, whereas holes flowing out of the lead contribute with charge $-e$, where $e$ is the electron charge. By definition, the transmission probability is related to the scattering matrix element through equation~\eqref{eq:transmission}.

The Fermi functions are defined as
\begin{equation}
    f_i(\epsilon)
    :=\frac{1}{1+\exp[\epsilon/(k_B\mathrm{T}_i)]},
    \label{eq:fermi}
\end{equation}
with $\mathrm{T}_i$ the temperature of lead $i$. We also define
\begin{equation}
    f_{i\text{e}}(\epsilon):=f_i(\epsilon-\mu_i),
    \qquad
    f_{i\text{h}}(\epsilon):=f_i(\epsilon+\mu_i).
\end{equation}
Since the superconducting chemical potential is chosen as $\mu_S=0$,
\begin{equation}
    f_{S\text{e}}(\epsilon)=f_{S\text{h}}(\epsilon)=f_S(\epsilon).
\end{equation}

Equation~\eqref{eq:charge_current} gives the outgoing charge current from the normal lead. Current conservation gives
\begin{equation}
    I_N+I_S=0.
\end{equation}
Therefore, there is only one independent current, which we define as
\begin{equation}
    I:=I_N,
    \qquad
    I_S=-I_N=-I.
\end{equation}

We define the Andreev transmission probability as
\begin{equation}
    T_A(\epsilon):=T_{NN}^{\text{he}}(\epsilon),
    \label{andreevtransmission}
\end{equation}
which can be interpreted as the probability for an electron to be reflected as a hole in the same normal contact, corresponding to Andreev reflection. Due to microreversibility and particle--hole symmetry, with the equations~\eqref{eq: microreversivility same contact} and~\eqref{eq:PHS}, then we obtain
\begin{equation}
    T_A(\epsilon)=T_A(-\epsilon),
    \qquad
    T_{NN}^{\text{eh}}(\epsilon)=T_{NN}^{\text{he}}(\epsilon)=T_A(\epsilon).
\end{equation}

Furthermore, we define the quasiparticle transmission probability as
\begin{equation}
    T_Q(\epsilon)
    :=T_{SN}^{\text{ee}}(\epsilon)+T_{SN}^{\text{he}}(\epsilon),
    \label{quasiparticletranmission}
\end{equation}
which can be interpreted as the probability for an electron to be transmitted as either an electron or a hole to the superconducting lead, corresponding to quasiparticle transmission. Due to microreversibility and particle--hole symmetry, with the equations~\eqref{eq:symmetric_scattering_matrix} and~\eqref{eq:PHS}, 
\begin{equation}
    T_{NS}^{\text{ee}}(\epsilon)+T_{NS}^{\text{eh}}(\epsilon)
    =T_{SN}^{\text{ee}}(\epsilon)+T_{SN}^{\text{he}}(\epsilon)
    =T_Q(\epsilon),
\end{equation}
\begin{equation}
    T_{NS}^{\text{hh}}(\epsilon)+T_{NS}^{\text{he}}(\epsilon)
    =T_{NS}^{\text{ee}}(-\epsilon)+T_{NS}^{\text{eh}}(-\epsilon)
    =T_Q(-\epsilon).
\end{equation}

With these definitions, using also the unitarity of the scattering matrix, expressed in the equation~\eqref{eq:unitarity_transmissions_explicit}, then
\begin{equation}
\begin{gathered}
    T_{NN}^{\text{ee}}(\epsilon)
    =1-T_{NN}^{\text{eh}}(\epsilon)
      -T_{NS}^{\text{ee}}(\epsilon)
      -T_{NS}^{\text{eh}}(\epsilon)
    =1-T_A(\epsilon)-T_Q(\epsilon),
    \\
    T_{NN}^{\text{hh}}(\epsilon)
    =T_{NN}^{\text{ee}}(-\epsilon)
    =1-T_A(-\epsilon)-T_Q(-\epsilon).
\end{gathered}
\end{equation}

We now expand Eq.~\eqref{eq:charge_current}, insert the corresponding $T_{ij}^{\alpha\beta}(\epsilon)$ in terms of $T_{A,Q}(\pm\epsilon)$, and make the change of integration variable $\epsilon\to-\epsilon$ in the appropriate terms. Using
\begin{equation}
    f_\alpha(-\epsilon)=1-f_\alpha(\epsilon),
\end{equation}
we obtain
\begin{equation}
    I
    =\frac{2e}{h}\int_{-\infty}^{\infty}\mathrm{d}\epsilon
    \left\{
    T_A\left(f_{N\text{e}}-f_{N\text{h}}\right)
    +T_Q\left(f_{N\text{e}}-f_S\right)
    \right\}.
    \label{eq:charge_current_TA_TQ}
\end{equation}
All terms are evaluated at $\epsilon$, for example $T_A\equiv T_A(\epsilon)$, $f_{N\text{e}}\equiv f_{N\text{e}}(\epsilon)$, etc. Equation~\eqref{eq:charge_current_TA_TQ} shows that the charge current can be split into Andreev and quasiparticle contributions,
\begin{equation}
    I=I_A+I_Q,
\end{equation}
with
\begin{equation}
    I_{A,Q}
    =\frac{2e}{h}\int_{-\infty}^{\infty}\mathrm{d}\epsilon\,
    T_{A,Q}\left(f_{N\text{e}}-f_{N\text{h},S}\right).
\end{equation}
Here the notation $f_{N\text{h},S}$ means $f_{N\text{h}}$ for the Andreev part and $f_S$ for the quasiparticle part.

\subsection{Energy and heat currents. Entropy production}
\label{sec:energy_heat_entropy}

The energy current~\cite{Buttiker1992,Blanter2000,Anantram1996} is given by
\begin{equation}
    J_N^E
    =\frac{1}{h}\int_{-\infty}^{\infty}\mathrm{d}\epsilon\,
    \epsilon\sum_{\alpha\beta j}
    \left[\delta_{Nj}\delta_{\alpha\beta}
    -T_{Nj}^{\alpha\beta}(\epsilon)\right]
    f_{j\beta}(\epsilon).
    \label{app:energy-current}
\end{equation}
This is the same expression as the charge current, but with $\epsilon$ instead of $\mathrm{sgn}(\alpha)e=\pm e$, because particles carry an energy $\epsilon$ contributing to the energy current, instead of a charge $\pm e$ contributing to the charge current. We choose the same sign convention as for charge current: energy flowing out of the lead is positive. Energy conservation imposes
\begin{equation}
    J_N^E+J_S^E=0.
\end{equation}
Thus, there is a single independent energy current,
\begin{equation}
    J^E:=J_N^E,
    \qquad
    J_S^E=-J_N^E=-J^E.
\end{equation}
Following the same procedure as for the charge current, the energy current can be expressed as
\begin{equation}
    J^E
    =\frac{2}{h}\int_{-\infty}^{\infty}\mathrm{d}\epsilon\,\epsilon
    \left\{
    T_A\left(f_{N\text{e}}-f_{N\text{h}}\right)
    +T_Q\left(f_{N\text{e}}-f_S\right)
    \right\}.
    \label{eq:energy_current_TA_TQ}
\end{equation}
All terms are evaluated at $\epsilon$, as in the charge-current expression. Equation~\eqref{eq:energy_current_TA_TQ} shows that the energy current can be split into Andreev and quasiparticle contributions,
\begin{equation}
    J^E=J_A^E+J_Q^E,
\end{equation}
with
\begin{equation}
    J_{A,Q}^{E}
    =\frac{2}{h}\int_{-\infty}^{\infty}\mathrm{d}\epsilon\,
    \epsilon\,T_{A,Q}\left(f_{N\text{e}}-f_{N\text{h},S}\right).
\end{equation}
Moreover, the integrand of $J_A^E$ is odd and is integrated over the symmetric interval $(-\infty,\infty)$, hence
\begin{equation}
    J_A^E=0.
\end{equation}
Therefore, the Andreev process does not contribute to the energy current and
\begin{equation}
    J^E=J_Q^E.
    \label{eq:JEequalJEQ}
\end{equation}

From the energy current, the heat current flowing out of each lead is defined as
\begin{equation}
    J_i :=J_i^E-\frac{\mu_i}{e}I_i.
    \label{eq:heat_current_definition}
\end{equation}
Therefore, the heat current can also be split into Andreev and quasiparticle parts,
\begin{equation}
    J_i=J_{i;A}+J_{i;Q},
\end{equation}
with
\begin{equation}
    J_{i;A,Q}=J_{i;A,Q}^{E}-\frac{\mu_i}{e}I_{i;A,Q}.
\end{equation}

The equation~\eqref{eq:heat_current_definition} comes from the first law of thermodynamics, where the change of total internal energy of the system $dU$ is attributed to the change of energy as work $\dbar W$ and the change of energy not as work, i.e, as heat, $\dbar Q$. The sign convention is taken as positive net work when it is performed by the system, so the energy as work is leaving the system and thus it subtracts from the internal energy as it leaves the system, and positive net heat when it is entering the system, so it sums yo the internal energy as the energy is going into the system. This way, we can write

\begin{equation}
    dU = \dbar Q - \dbar W,
    \label{1lawtermo}
\end{equation}
where the $\dbar$ means an inexact differential, since heat and work depend on the path and not only on the final and initial states. We see that if $\dbar W$ is positive, then the internal energy is reduced as it is work performed by the system by convention, so energy leaving the system, and if it is negative, the system receive a work which increases its internal energy. Furthermore, if $\dbar Q$ is positive, then the heat is entering the system, so increases its internal energy, and vice versa if it is negative. Then the energy current is
\begin{equation}
    J^E = J - J^W.
    \label{1lawtermonoeq}
\end{equation}

Informally, we have just divided by a time interval $dt$ the equation~\eqref{1lawtermo} and thus taking the change in time (currents). The change of energy of the system due to an exchange of particles is the change of particles $I^{\text{particles}}$ times the change of energy with a change of the number of particles (which is the chemical potential $\mu$), i.e., the number of particles flowing times the energy per particle flowing, $\mu I^{\text{particles}}$. As the work is performed only by an exchange of particles, and the work sign is opposite to the sign of the change of internal energy (positive work performed means less internal energy, as explained above), then $J^W = - \mu I^{\text{particles}}$. Informally, we can take $J^W = \frac{\dbar W}{dt} = \frac{\dbar W}{dN_{\text{particles}}}\frac{dN_{\text{particles}}}{dt} = - \frac{\dbar U}{dN_{\text{particles}}}\frac{dN_{\text{particles}}}{dt} = - \mu I^{\text{particles}}$, where $N_{\text{particles}}$ is the number of particles. Finally, as the charge current is the number of particles flowing times the charge per particle, and every particle has the same charge $e$, then $I = e I^{\text{particles}}$, so $I^{\text{particles}} = \frac{I}{e}$. Thus, we have that $J^W = - \frac{\mu}{e}I$. Inserting this into equation~\eqref{1lawtermonoeq}, we obtain the equation~\eqref{eq:heat_current_definition}.

The entropy production rate~\cite{Seifert2012,Horowitz2020} is defined as
\begin{equation}
    \sigma
    :=-\sum_i\frac{J_i}{\mathrm{T}_i}
    =-\frac{J_N}{\mathrm{T}_N}-\frac{J_S}{\mathrm{T}_S},
    \label{eq:entropy_production_definition}
\end{equation}
which comes from the second law of thermodynamic and the Clausius formula for entropy as $- \sum_i \frac{\dbar Q_i}{T_i}$, informally dividing by a time interval $dt$, i.e., differentiating with respect to time, and thus taking the change in time (currents). It can also be split into Andreev and quasiparticle parts,
\begin{equation}
    \sigma=\sigma_A+\sigma_Q,
\end{equation}
with
\begin{equation}
    \sigma_{A,Q}=-\sum_i\frac{J_{i;A,Q}}{\mathrm{T}_i}.
\end{equation}

Since $J_N^E=J^E=J_Q^E$, as shown in~\eqref{eq:JEequalJEQ}, from Eq.~\eqref{eq:heat_current_definition} we can write explicitly
\begin{equation}
\begin{gathered}
    J_N
    =J_Q^E-\frac{\mu_N}{e}I_N
    =\left(J_Q^E-\frac{\mu_N}{e}I_Q\right)
     +\left(-\frac{\mu_N}{e}I_A\right),
    \\
   \implies J_{N;Q}=J_Q^E-\frac{\mu_N}{e}I_Q,
    \qquad
    J_{N;A}=-\frac{\mu_N}{e}I_A.
\end{gathered}
\end{equation}
Furthermore, since $\mu_S = 0$, then from Eq.~\eqref{eq:heat_current_definition} and using the Eq.~\eqref{eq:JEequalJEQ} we get $J_S = J_S^E=-J^E=-J_Q^E$, so
\begin{equation}
    J_{S;Q}=-J_Q^E,
    \qquad
    J_{S;A}=0.
\end{equation}
Thus, the Andreev and quasiparticle contributions to the entropy production rate are
\begin{equation}
    \begin{gathered}
        \sigma_Q
        =-\left(\frac{J_Q^E-\frac{\mu_N}{e}I_Q}{\mathrm{T}_N}\right)
         -\left(\frac{-J_Q^E}{\mathrm{T}_S}\right),
        \\
        \sigma_A
        =\frac{\mu_N}{e\mathrm{T}_N}I_A.
    \end{gathered}
    \label{eq:sigma_A_Q}
\end{equation}

\subsection{Charge noise}
\label{sec:charge_noise}

We now compute the charge-current noise, which corresponds (by definition) to the symmetrized covariance
\begin{equation}
\begin{gathered}
    \mathcal{S}_{ij}(t,t_0)
    :=\frac{1}{2}\left\langle
    \Delta\hat{I}_i(t)\Delta\hat{I}_j(t_0)
    +\Delta\hat{I}_j(t)\Delta\hat{I}_i(t_0)
    \right\rangle,
    \\
    \mathcal{S}_{ij}(t,t_0)=\mathcal{S}_{ij}(t-t_0),
    \\
    \Delta\hat{I}_i=\hat{I}_i-\langle\hat{I}_i\rangle.
    \label{eq:noisedefinition}
\end{gathered}
\end{equation}
This is equivalent to taking directly $t_0=0$ and $t$ as the variable, as the noise only depends on the time lapse $\tau : = t-t_0$, which follows from assuming a nonequilibrium steady state (NESS). We can obtain the noise in frequency domain by computing the Fourier transformation of the Eq.~\eqref{eq:noisedefinition} as follows
\begin{equation}
    \mathcal{S}_{ij}(\omega)\,2\pi\delta(\omega+\omega_0)
    =\left\langle
    \frac{1}{2}\left[
    \Delta\hat{I}_i(\omega)\Delta\hat{I}_j(\omega_0)
    +\Delta\hat{I}_j(\omega)\Delta\hat{I}_i(\omega_0)
    \right]
    \right\rangle,
\end{equation}
with $\hat{I}_i(\omega)$ the Fourier transform of $\hat{I}_i(t)$. Using the scattering formula for the zero-frequency noise ($\omega=0$)~\cite{Anantram1996,Blanter2000}, we obtain
\begin{equation}
    \mathcal{S}_{NN}
    :=\mathcal{S}_{NN}(\omega=0)
    =\mathcal{S}_{NN}^{AA}+\mathcal{S}_{NN}^{AB},
\end{equation}
with
\begin{gather}
\begin{gathered}
\mathcal{S}_{NN}^{AA}
=
\frac{e^2}{h}
\int_{-\infty}^{\infty}\mathrm{d}\epsilon
\sum_{\alpha}
\left\{
\left[1-T_{NN}^{\alpha\alpha}(\epsilon)\right]^2
f_{N\alpha}(\epsilon)
\left[1-f_{N\alpha}(\epsilon)\right]
\right.
\left.
+
\sum_{(k\gamma,l\delta)\neq(N\alpha,N\alpha)}
T_{Nk}^{\alpha\gamma}(\epsilon)
T_{Nl}^{\alpha\delta}(\epsilon)
 f_{k\gamma}(\epsilon)
\left[1-f_{l\delta}(\epsilon)\right]
\right\},
\end{gathered}
\label{eq:S_AA_NN}
\\[1em]
\begin{gathered}
\mathcal{S}_{NN}^{AB}
=
\frac{e^2}{h}
\int_{-\infty}^{\infty}\mathrm{d}\epsilon
\sum_{\alpha}
\left\{
2T_{NN}^{\alpha\bar{\alpha}}(\epsilon)
 f_{N\bar{\alpha}}(\epsilon)
\left[1-f_{N\bar{\alpha}}(\epsilon)\right]
\right.
\left.
+
\sum_{k\gamma}
 s_{Nk}^{\bar{\alpha}\gamma}(\epsilon)
 s_{Nk}^{\alpha\gamma\dagger}(\epsilon)
 f_{k\gamma}(\epsilon)
\sum_{l\delta}
 s_{Nl}^{\alpha\delta}(\epsilon)
 s_{Nl}^{\bar{\alpha}\delta\dagger}(\epsilon)
 f_{l\delta}(\epsilon)
\right\}.
\end{gathered}
\label{eq:S_AB_NN}
\end{gather}
Here
\begin{equation}
\bar{\alpha}=\begin{cases}
\text{h}, & \alpha=\text{e},\\
\text{e}, & \alpha=\text{h}.
\end{cases}
\end{equation}
Since $I_S=-I_N$,
\begin{equation}
    \mathcal{S}_{SS}=\mathcal{S}_{NN}.
\end{equation}
Thus, in the following we use
\begin{equation}
    \mathcal{S}:=\mathcal{S}_{NN}.
\end{equation}

For simplicity, we will usually refer to zero-frequency noise simply as just noise. Using the same procedure as with the charge and heat currents (sections~\ref{sec:charge_current} and~\ref{sec:energy_heat_entropy}) and using
\begin{equation}
    |ab|^2=|a|^2|b|^2,
\end{equation}
we obtain a decomposition of $\mathcal{S}$ into an Andreev part, a quasiparticle part, and a cross-correlation part involving interference terms:
\begin{equation}
    \mathcal{S}=\mathcal{S}_A+\mathcal{S}_Q+\mathcal{S}_{\mathrm{Cross}}.
\end{equation}
The explicit contributions are
\begin{gather}
\begin{gathered}
\mathcal{S}_{A}
=
\frac{4e^{2}}{h}
\int_{-\infty}^{\infty}\mathrm{d}\epsilon
\left\{
T_{A}
\left[
 f_{N\text{e}}(1-f_{N\text{e}})
+
 f_{N\text{h}}(1-f_{N\text{h}})
\right]
\right.
\\
\left.
+
T_{A}(1-T_{A})
(f_{N\text{e}}-f_{N\text{h}})^{2}
\right\},
\end{gathered}
\label{app:andreev-noise}
\\[0.8em]
\begin{gathered}
\mathcal{S}_{Q}
=
\frac{2e^{2}}{h}
\int_{-\infty}^{\infty}\mathrm{d}\epsilon
\left\{
T_{Q}
\left[
 f_{N\text{e}}(1-f_{N\text{e}})
+
 f_{S}(1-f_{S})
\right]
\right.
\\
\left.
+
T_{Q}(1-T_{Q})
(f_{N\text{e}}-f_{S})^{2}
\right\},
\end{gathered}
\label{app:quasiparticle-noise}
\\[1em]
\begin{gathered}
\mathcal{S}_{\mathrm{Cross}}
=
\frac{e^{2}}{h}
\int_{-\infty}^{\infty}\mathrm{d}\epsilon
\left\{
-2T_{A}T_{Q}
\big[
(f_{N\text{e}}-f_{N\text{h}})^{2}
+2(f_{N\text{e}}-f_{N\text{h}})(f_{N\text{e}}-f_{S})
\big]
\right.
\\
+
2
\left|
 s_{NN}^{\text{he}}s_{NN}^{\text{ee}\dagger}
+
 s_{NN}^{\text{hh}}s_{NN}^{\text{eh}\dagger}
\right|^{2}
(f_{N\text{h}}-f_{S})^{2}
\\
+
4\operatorname{Re}
\left[
 s_{NN}^{\text{he}}s_{NN}^{\text{ee}\dagger}
\left(
 s_{NN}^{\text{he}}s_{NN}^{\text{ee}\dagger}
+
 s_{NN}^{\text{hh}}s_{NN}^{\text{eh}\dagger}
\right)^{\dagger}
\right]
\left.
\times
(f_{N\text{e}}-f_{N\text{h}})(f_{N\text{h}}-f_{S})
\right\}.
\end{gathered}
\label{app:cross-noise}
\end{gather}

All terms are evaluated at $\epsilon$, as in the case of the charge current. This form is convenient because it makes transparent which terms cancel in the limits $\Delta\to0$~\cite{Blonder1982,Beenakker1997} and $\Delta\to\infty$~\cite{Andreev1964,Blonder1982,Beenakker1992,Governale2026}. In the first case, there are no processes involving electron--hole conversion, since the superconducting gap is zero. Thus, there are no Cooper pairs that can absorb electrons and emit holes, or vice versa, resulting in
\begin{equation}
    s_{NN}^{\text{he}}=s_{NN}^{\text{eh}}=0,
    \qquad
    T_A=0.
\end{equation}
Consequently,
\begin{equation}
    \mathcal{S}\to\mathcal{S}_Q
    \qquad
    \text{when }\Delta\to0,
\end{equation}
i.e., the N--S conductor becomes an N--N conductor.

In the opposite limit, $\Delta\to\infty$, there is no transmission to the superconducting lead as single electrons or holes, since all particles form Cooper pairs and all single particles $\alpha$ are reflected to the normal lead as the opposite particle $\bar\alpha$. Therefore,
\begin{equation}
    s_{NS}^{\alpha\beta}=0,
    \qquad
    T_Q=0.
\end{equation}
By unitarity of the scattering matrix,
\begin{equation}
    s_{NN}^{\text{he}}s_{NN}^{\text{ee}\dagger}
    +s_{NN}^{\text{hh}}s_{NN}^{\text{eh}\dagger}=0.
\end{equation}
Hence
\begin{equation}
    \mathcal{S}\to\mathcal{S}_A
    \qquad
    \text{when } \Delta\to\infty,
\end{equation}
i.e., only Andreev processes remain, as particles in the superconductor are always within the gap.

\subsection{Positivity of shot noise}
\label{sec:positivity_shot_noise}

We define the thermal noise as the noise remaining at equilibrium,
\begin{equation}
    \mathcal{S}^{\mathrm{th}}
    :=\mathcal{S}\rvert_{\mathrm{eq}},
\end{equation}
and the shot noise as
\begin{equation}
    \mathcal{S}^{\mathrm{sh}}
    :=\mathcal{S}-\mathcal{S}^{\mathrm{th}},
\end{equation}
so the total noise is the sum of both~\cite{Blanter2000}. For each contribution,
\begin{equation}
    \mathcal{S}_{A,Q,\mathrm{Cross}}^{\mathrm{th}}
    :=\mathcal{S}_{A,Q,\mathrm{Cross}}\rvert_{\mathrm{eq}},
\end{equation}
\begin{equation}
    \mathcal{S}_{A,Q,\mathrm{Cross}}^{\mathrm{sh}}
    =\mathcal{S}_{A,Q,\mathrm{Cross}}
    -\mathcal{S}_{A,Q,\mathrm{Cross}}^{\mathrm{th}}.
\end{equation}
The cross thermal contribution vanishes,
\begin{equation}
    \mathcal{S}_{\mathrm{Cross}}^{\mathrm{th}}=0,
\end{equation}
because all terms contain differences of the form $f_i-f_j$, which are zero at equilibrium:
\begin{equation}
    f_i\rvert_{\mathrm{eq}}
    =f_j\rvert_{\mathrm{eq}}
    =f_{\mathrm{eq}}.
\end{equation}
Therefore,
\begin{equation}
    \mathcal{S}^{\mathrm{th}}
    =\mathcal{S}_{A}^{\mathrm{th}}+
     \mathcal{S}_{Q}^{\mathrm{th}},
\end{equation}
\begin{equation}
    \mathcal{S}^{\mathrm{sh}}
    =\mathcal{S}_{A}^{\mathrm{sh}}
    +\mathcal{S}_{Q}^{\mathrm{sh}}
    +\mathcal{S}_{\mathrm{Cross}}.
\end{equation}
We can compute explicitly the thermal and shot contributions of the Andreev and quasiparticle sectors, which can be write as
\begin{equation}
\begin{gathered}
\mathcal{S}_{A,Q}^{\mathrm{th}}
=g_{A,Q}\frac{e^{2}}{h}
\int_{-\infty}^{\infty}\mathrm{d}\epsilon\,
T_{A,Q}
\left[
 f_{N\text{e}}(1-f_{N\text{e}})
 +f_{N\text{h},S}(1-f_{N\text{h},S})
\right],
\\
\mathcal{S}_{A,Q}^{\mathrm{sh}}
=g_{A,Q}\frac{e^{2}}{h}
\int_{-\infty}^{\infty}\mathrm{d}\epsilon\,
T_{A,Q}(1-T_{A,Q})
\left(f_{N\text{e}}-f_{N\text{h},S}\right)^2,
\end{gathered}
\label{eq:Sth_Ssh_AQ}
\end{equation}
with
\begin{equation}
    g_A=4,
    \qquad
    g_Q=2.
\end{equation}

After some algebra, we introduce the shorthand notation
\begin{equation}
\begin{gathered}
    x=f_{N\text{e}}-f_{N\text{h}},
    \qquad
    y=f_{N\text{e}}-f_S,
    \qquad
    z=y-x=f_{N\text{h}}-f_S,
    \\
    a=s_{NN}^{\text{he}}s_{NN}^{\text{ee}\dagger},
    \qquad
    b=s_{NN}^{\text{hh}}s_{NN}^{\text{eh}\dagger},
    \qquad
    c=a+b.
\end{gathered}
\end{equation}
Then the shot noise can be written as
\begin{equation}
\begin{gathered}
\mathcal{S}^{\mathrm{sh}}
=
\frac{2e^{2}}{h}
\int_{-\infty}^{\infty}\mathrm{d}\epsilon
\left\{
\left|cz+ax\right|^{2}
+
T_{Q}(1-T_{Q})
\left(
 y-\frac{T_{A}}{1-T_{Q}}x
\right)^{2}
\right.
\\
\left.
+
\frac{T_{A}(1-T_{A}-T_{Q})}{1-T_{Q}}
 x^{2}
\right\}.
\end{gathered}
\label{eq:Ssh_positive}
\end{equation}
We note that
\begin{equation}
    0\leq |s_{NN}^{\text{he}}|^2 = T_{A}\leq1, 
    \qquad
    0\leq |s_{NN}^{\text{ee}}|^2=1-T_A-T_Q\leq1, 
    \qquad
    \implies 0\leq  T_{Q}\leq1 .
    \label{eq: POSITIVITY TRANSMISSIONS}
\end{equation}
Equations~\eqref{eq:Ssh_positive} and~\eqref{eq: POSITIVITY TRANSMISSIONS} imply
\begin{equation}
    \mathcal{S}^{\mathrm{sh}}\geq0.
\end{equation}
Therefore, in hybrid normal--superconducting devices,
\begin{equation}
    \mathcal{S}
    =\mathcal{S}^{\mathrm{th}}+
     \mathcal{S}^{\mathrm{sh}}
    \geq\mathcal{S}^{\mathrm{th}}
    =\mathcal{S}_{A}^{\mathrm{th}}+
     \mathcal{S}_{Q}^{\mathrm{th}}.
    \label{eq:S_geq_Sth}
\end{equation}
Also,
\begin{equation}
    \mathcal{S}_{A,Q}^{\mathrm{sh}}\geq0,
\end{equation}
because $0\leq T_{A,Q}\leq1$. Indeed, $T_A$ is the squared modulus of an element of the scattering matrix and $1-T_A-T_Q$ goes from $0$ to $1$, with $T_A$ also going from $0$ to $1$, so $T_Q$ must also lie between $0$ and $1$.

Finally, since
\begin{equation}
    0\leq f_i\leq1,
\end{equation}
we also have
\begin{equation}
    \mathcal{S}_{A,Q}^{\mathrm{th}}\geq0,
    \qquad
    \mathcal{S}^{\mathrm{th}}\geq0.
\end{equation}

\section{Hybrid quantum thermodynamic uncertainty relation}
\label{sec:HQTUR}

Following the steps of Brandner and Saito~\cite{Brandner2025,Brandner2018}, for the quasiparticle part we obtain
\begin{equation}
    \sigma_Q
    \geq
    \frac{2k_B|I_Q|}{e}\,
    \operatorname{arsinh}\!\left(
    \frac{e|I_Q|}{\mathcal{S}_{Q}^{\mathrm{th}}}
    \right)
    \geq
    \frac{k_B|I_Q|}{e}\,
    \operatorname{arsinh}\!\left(
    \frac{2e|I_Q|}{\mathcal{S}_{Q}^{\mathrm{th}}}
    \right).
\label{eq:HQTUR_Q}
\end{equation}
In the last step we used
\begin{equation}
    2\operatorname{arsinh}(x)\geq\operatorname{arsinh}(2x),
    \qquad \forall x\geq0.
\end{equation}
and $\frac{e|I_Q|}{\mathcal{S}_{Q}^{\mathrm{th}}} \geq 0$ since the quasiparticle thermal noise is non-negative. For the Andreev part, following the same steps, we recover the result of Mayo \emph{et al.}~\cite{Governale2026},
\begin{equation}
    \sigma_A
    \geq
    \frac{k_B|I_A|}{e}\,
    \operatorname{arsinh}\!\left(
    \frac{2e|I_A|}{\mathcal{S}_{A}^{\mathrm{th}}}
    \right).
    \label{eq:HQTUR_A}
\end{equation}
where $\frac{2e|I_A|}{\mathcal{S}_{A}^{\mathrm{th}}} \geq 0$ as the Andreev thermal noise is also non-negative. Furthermore, every quantity entering these inequalities is non-negative. Summing equations~\eqref{eq:HQTUR_Q} and~\eqref{eq:HQTUR_A} gives
\begin{equation}
    \sigma = \sigma_A+\sigma_Q
    \geq
    \frac{k_B|I_Q|}{e}\,
    \operatorname{arsinh}\!\left(
    \frac{2e|I_Q|}{\mathcal{S}_{Q}^{\mathrm{th}}}
    \right)
    +
    \frac{k_B|I_A|}{e}\,
    \operatorname{arsinh}\!\left(
    \frac{2e|I_A|}{\mathcal{S}_{A}^{\mathrm{th}}}
    \right).
\label{eq:HQTUR_sum}
\end{equation}

We define the function
\begin{equation}
\begin{aligned}
\phi \colon \mathbb{R} &\to \mathbb{R}, \\
x &\mapsto \phi(x) := x \, \operatorname{arsinh}\!\left(2e x \right),
\end{aligned}
\label{eq:phi(x)}
\end{equation}
whose second derivative with respect to \(x\) reads
\begin{equation}
    \phi^{\prime\prime}(x)=\frac{4e(1+2e^2x^2)}{(1+4e^2x^2)^{3/2}}.
\end{equation}
Since the elementary charge $e$ is a positive constant, it follows that
\begin{equation}
    \phi^{\prime\prime}(x) > 0 \qquad \forall x \in \mathbb{R},
\end{equation}
which implies that the function \(\phi\) is convex, more precisely, strictly convex. Hence, by definition of a convex function, for all $t \in [0,1]$ and all \(x_1,x_2\in\operatorname{dom}(\phi)=\mathbb{R}\), we have
\begin{equation}
 t \, \phi(x_1) + \left(1-t \right) \phi(x_2) \geq \phi \!\left (t \, x_1 + \left(1-t \right) x_2 \right),
 \label{convexity}
\end{equation}
where \(\operatorname{dom}(\phi)\) denotes the domain of \(\phi\). Now, we choose
\begin{equation}
    t : = \frac{\mathcal{S}^{th}_A}{\mathcal{S}^{th}_A + \mathcal{S}^{th}_Q}, \qquad 1-t = \frac{\mathcal{S}^{th}_Q}{\mathcal{S}^{th}_A + \mathcal{S}^{th}_Q}, \qquad x_1 := \frac{|I_A|}{\mathcal{S}^{th}_A}, \qquad x_2 := \frac{|I_Q|}{\mathcal{S}^{th}_Q}.
    \label{eq:definitionstx1x2}
\end{equation}
Since $\mathcal{S}_{A}^{\mathrm{th}}+\mathcal{S}_{Q}^{\mathrm{th}} \geq\mathcal{S}_{A,Q}^{\mathrm{th}}\geq0$, then $0 \leq t \leq 1$, and $x_{1,2} \geq 0$. Substituting this $t$ in equation~\eqref{convexity}, we get that
\begin{equation}
  \mathcal{S}^{th}_A \phi(x_1) + \mathcal{S}^{th}_Q \phi(x_2) \geq \left(\mathcal{S}^{th}_A + \mathcal{S}^{th}_Q \right)\phi \!\left (\frac{\mathcal{S}^{th}_A x_1 + \mathcal{S}^{th}_Q x_2}{\mathcal{S}^{th}_A + \mathcal{S}^{th}_Q}\right).
\end{equation}
Substituting $\phi(x)$ from equation~\eqref{eq:phi(x)}, we obtain
\begin{equation}
  \mathcal{S}^{th}_A x_1 \operatorname{arsinh}\!\left( 2e x_1 \right) + \mathcal{S}^{th}_Q x_2 \operatorname{arsinh}\!\left( 2e x_2 \right) \geq \left( \mathcal{S}^{th}_A x_1 + \mathcal{S}^{th}_Q x_2 \right)\operatorname{arsinh}\! \left (2e\frac{\mathcal{S}^{th}_A x_1 + \mathcal{S}^{th}_Q x_2}{\mathcal{S}^{th}_A + \mathcal{S}^{th}_Q}\right),
\end{equation}
and substituting $x_1$ and $x_2$ from equation~\eqref{eq:definitionstx1x2}, then
\begin{equation}
  |I_A| \operatorname{arsinh}\!\left(\frac{ 2e |I_A|}{\mathcal{S}^{th}_A} \right) + |I_Q|  \operatorname{arsinh}\!\left( \frac{2e|I_Q|}{\mathcal{S}^{th}_Q} \right) \geq \left(|I_A| + |I_Q| \right) \operatorname{arsinh}\! \left (2e\frac{|I_A| + |I_Q| }{\mathcal{S}^{th}_A + \mathcal{S}^{th}_Q}\right),
  \label{jensenineq}
\end{equation}
We then use the following properties:
\begin{alignat}{2}
  &\text{Triangle inequality:}
    &\qquad& |I_A|+|I_Q|\geq|I_A+I_Q| , \\
  &\text{Monotonicity of }\operatorname{arsinh}:
    &\qquad& \operatorname{arsinh}(x)\geq\operatorname{arsinh}(y)
    \qquad \text{for all }x\geq y\geq0.
    \label{monton arcsinh}
\end{alignat}
Substituting this in the equation~\eqref{jensenineq} and multiplying by $k_B/e$, we obtain
\begin{equation}
\frac{k_B \left(|I_A| + |I_Q| \right)}{e} \operatorname{arsinh}\! \left (2e\frac{|I_A| + |I_Q| }{\mathcal{S}^{th}_A + \mathcal{S}^{th}_Q}\right) \geq\frac{k_B|I_A+I_Q|}{e} \operatorname{arsinh}\! \left (2e\frac{|I_A + I_Q| }{\mathcal{S}^{th}_A + \mathcal{S}^{th}_Q}\right)
\end{equation}
\begin{equation}
 \implies \frac{k_B|I_A|}{e} \operatorname{arsinh}\!\left(\frac{ 2e |I_A|}{\mathcal{S}^{th}_A} \right) + \frac{k_B|I_Q|}{e}  \operatorname{arsinh}\!\left( \frac{2e|I_Q|}{\mathcal{S}^{th}_Q} \right) \geq\frac{k_B|I_A+I_Q|}{e} \operatorname{arsinh}\! \left (2e\frac{|I_A + I_Q| }{\mathcal{S}^{th}_A + \mathcal{S}^{th}_Q}\right).
\end{equation}
Thus,
\begin{equation}
\begin{gathered}
    \sigma
    \geq
    \frac{k_B|I_A+I_Q|}{e}\,
    \operatorname{arsinh}\!\left(
    \frac{2e|I_A+I_Q|}
    {\mathcal{S}_{A}^{\mathrm{th}}+\mathcal{S}_{Q}^{\mathrm{th}}}
    \right)
    \\
    =
    \frac{k_B|I|}{e}\,
    \operatorname{arsinh}\!\left(
    \frac{2e|I|}{\mathcal{S}^{\mathrm{th}}}
    \right),
\end{gathered}
\label{eq:HQTUR_Sth}
\end{equation}
with $I = I_A + I_Q$ and $\mathcal{S}^{\mathrm{th}} = \mathcal{S}^{\mathrm{th}}_A + \mathcal{S}^{\mathrm{th}}_Q$. Finally, from equation~\eqref{eq:S_geq_Sth},
\begin{equation}
    \mathcal{S}\geq\mathcal{S}^{\mathrm{th}}\geq0.
\end{equation}
Therefore,
\begin{equation}
    \frac{1}{\mathcal{S}^{\mathrm{th}}}
    \geq
    \frac{1}{\mathcal{S}}
    \geq0.
\end{equation}
Using again the monotonicity of $\operatorname{arsinh}(x)$ for all $x\geq0$, i.e, the equation~\eqref{monton arcsinh}, we obtain the HQTUR
\begin{equation}
    \sigma
    \geq
    \frac{k_B|I|}{e}\,
    \operatorname{arsinh}\!\left(
    \frac{2e|I|}{\mathcal{S}}
    \right).
    \label{eq:HQTUR_final}
\end{equation}
This proves that the HQTUR, previously established in the pure-Andreev
limit \(\Delta\to\infty\), remains valid for an arbitrary real
superconducting gap \(\Delta\).

\section{Equivalence between the scattering-matrix and Green's-function formalisms}
\label{app:green-functions}
In this section, we establish the connection between the scattering-matrix approach and the Green's-function formalism. This equivalence is particularly useful for the configurations considered throughout the figures, where the central conductor is modeled as a quantum dot. In practice, the scattering matrix is obtained from the retarded Green's function of the conductor through the Fisher--Lee relation, allowing us to compute the relevant scattering amplitudes directly from the Green's-function description.

\subsection{Green's functions and Fisher--Lee relation in hybrid Normal--Superconducting systems}
Our Hamiltonian can be written as
\begin{equation}
    H = H_N + H_S + H_C + H_{T},
    \label{hamiltoniancomplet}
\end{equation}
where $H_N$ is the Hamiltonian of the normal lead, $H_S$ is the Hamiltonian of the superconducting lead, $H_C$ is the Hamiltonian of the central conductor (scatter), and $H_{T}$ is the coupling Hamiltonian between leads and the conductor~\cite{Blonder1982,Beenakker1997,deFranceschi2010}.

The normal lead is described by a non-interacting fermionic Hamiltonian
\begin{equation}
    H_N
    =
    \sum_{k,\sigma}
    \varepsilon_{Nk}
    c^{\dagger}_{Nk\sigma}
    c_{Nk\sigma},
\end{equation}
where $c^{\dagger}_{Nk\sigma}$ creates an electron in the normal lead with momentum $k$ and spin $\sigma$, and $\varepsilon_{Nk}$ is the single-particle energy in the normal lead.

The superconducting lead is described by a mean-field BCS Hamiltonian~\cite{BardeenCooperSchrieffer1957,deGennes1966},
\begin{equation}
    H_S
    =
    \sum_{k,\sigma}
    \varepsilon_{Sk}
    c^{\dagger}_{Sk\sigma}
    c_{Sk\sigma}
    -
    \sum_k
    \left(
    \Delta
    c^{\dagger}_{Sk\uparrow}
    c^{\dagger}_{S,-k\downarrow}
    +
    \Delta^{*}
    c_{S,-k\downarrow}
    c_{Sk\uparrow}
    \right) +\mathrm{const},
    \label{Superconducting Hamiltonian equation}
\end{equation}
where
\begin{equation}
    \Delta
    =
    |\Delta|e^{i\phi_S}.
\end{equation}
We are going to take $\Delta^* = \Delta$. The first term describes the normal part of the superconducting lead, while the second term describes the creation and annihilation of Cooper pairs.

The central conductor is kept general and can be written as
\begin{equation}
    H_C
    =
    \sum_{i,j,\sigma}
    h_{ij}
    d^{\dagger}_{i\sigma}
    d_{j\sigma}
    +
    H_{\mathrm{int}},
    \label{Hconductor}
\end{equation}
where $d^{\dagger}_{i\sigma}$ creates an electron in the central conductor, $h_{ij}$ contains the single-particle terms of the conductor, and $H_{\mathrm{int}}$ accounts for possible interaction terms. In a non-interacting conductor, one simply takes $H_{\mathrm{int}}=0$.

The coupling between the leads and the central conductor is described by the tunneling Hamiltonian
\begin{equation}
    H_T
    =
    \sum_{\ell = N,S}
    \sum_{k,i,\sigma}
    \left(
    t_{\ell k i}
    c^{\dagger}_{\ell k\sigma}
    d_{i\sigma}
    +
    t_{\ell k i}^{*}
    d^{\dagger}_{i\sigma}
    c_{\ell k\sigma}
    \right),
\end{equation}
where $\ell=N,S$ labels the normal and superconducting leads, respectively, and $t_{\ell k i}$ is the tunneling amplitude between state $(\ell,k,\sigma)$ in the lead and state $(i,\sigma)$ in the central conductor.

Using the electron-hole basis introduced above, it is convenient to define the Nambu spinors
\begin{equation}
    \bm{c}_{\ell k\sigma}
    :=
    \begin{pmatrix}
        c_{\ell k\sigma}\\
        c^{\dagger}_{\ell,-k\bar{\sigma}}
    \end{pmatrix},
    \qquad
    \bm{d}_{i\sigma}
    :=
    \begin{pmatrix}
        d_{i\sigma}\\
        d^{\dagger}_{i\bar{\sigma}}
    \end{pmatrix}.
\end{equation}
In this basis, the normal lead Hamiltonian can be written, up to an irrelevant constant, as
\begin{equation}
    H_N
    =
    \frac{1}{2}
    \sum_{k,\sigma}
    \bm{c}^{\dagger}_{Nk\sigma}
    \left(
    \varepsilon_{Nk}\tau_z
    \right)
    \bm{c}_{Nk\sigma},
\end{equation}
where $\tau_z$ is the Pauli matrix acting in electron-hole space,
\begin{equation}
    \tau_z
    =
    \begin{pmatrix}
        1 & 0\\
        0 & -1
    \end{pmatrix}.
\end{equation}
The factor $1/2$ appears because the Nambu basis doubles the degrees of freedom by treating electron and hole components simultaneously, and therefore avoids double counting.

Analogously, the superconducting lead Hamiltonian can be written as
\begin{equation}
    H_S
    =
    \sum_k
    \bm{c}^{\dagger}_{Sk\uparrow}
    \begin{pmatrix}
        \varepsilon_{Sk} & -\Delta\\
        -\Delta^{*} & -\varepsilon_{Sk}
    \end{pmatrix}
    \bm{c}_{Sk\uparrow} +\mathrm{const}.
\end{equation}

The conductor Hamiltonian can also be written in Nambu form~\cite{Bogoliubov1958,deGennes1966} as
\begin{equation}
    H_C
    =
    \frac{1}{2} \sum_{ij\sigma}
    \bm{d}^{\dagger}_{i\sigma}
    \left(\bm{H}_C^{\mathrm{BdG}}\right)_{ij}
    \bm{d}_{j\sigma} = \frac{1}{2} \sum_{\sigma}
    \bm{d}^{\dagger}_{\sigma}
    \bm{H}_C^{\mathrm{BdG}}
    \bm{d}_{\sigma},
\end{equation}
with $\bm{H}_C^{\mathrm{BdG}}$ is the Bogoliubov--de Gennes Hamiltonian of the central region, and $\bm{d}_\sigma$ is a block vector, where each block component $i$ is a vector $\bm{d}_{i\sigma}$, i.e.
\begin{equation}
    \bm{d}_\sigma := \begin{pmatrix}
        \bm{d}_{1\sigma} \\ \bm{d}_{2\sigma} \\ \vdots \\ \bm{d}_{n\sigma} 
    \end{pmatrix} = \begin{pmatrix}
            \begin{pmatrix}
        d_{1\sigma}\\
        d^{\dagger}_{1\bar{\sigma}}
    \end{pmatrix} \\     \begin{pmatrix}
        d_{2\sigma}\\
        d^{\dagger}_{2\bar{\sigma}}
    \end{pmatrix} \\ \vdots \\     \begin{pmatrix}
        d_{n\sigma}\\
        d^{\dagger}_{n\bar{\sigma}}
    \end{pmatrix} 
    \end{pmatrix} \equiv \begin{pmatrix}
        d_{1\sigma}\\
        d^{\dagger}_{1\bar{\sigma}} \\
        d_{2\sigma}\\
        d^{\dagger}_{2\bar{\sigma}}
\\ \vdots \\     
        d_{n\sigma}\\
        d^{\dagger}_{n\bar{\sigma}}
    \end{pmatrix} 
\end{equation}

In the absence of intrinsic superconductivity in the conductor, this matrix takes the block form
\begin{equation}
    \bm{H}_C^{\mathrm{BdG}}
    =
    \begin{pmatrix}
        \bm{h}_C & 0\\
        0 & -\bm{h}_C^{*}
    \end{pmatrix},
\end{equation}
where $\bm{h}_C$ is the matrix whose elements are $h_{ij}$ of the equation~\eqref{Hconductor}. Finally, the tunneling Hamiltonian can be expressed in Nambu space as
\begin{equation}
    H_T
    =
    \frac{1}{2}
    \sum_{\ell=N,S}
    \sum_{k,i,\sigma}
    \left(
    \bm{c}^{\dagger}_{\ell k\sigma}
    \bm{t}_{\ell k i}
    \bm{d}_{i\sigma}
    +
    \bm{d}^{\dagger}_{i\sigma}
    \bm{t}^{\dagger}_{\ell k i}
    \bm{c}_{\ell k\sigma}
    \right),
\end{equation}
where, for spin-conserving tunneling, one may choose
\begin{equation}
    \bm{t}_{\ell k i}
    :=
    \begin{pmatrix}
        t_{\ell k i} & 0\\
        0 & -t^{*}_{\ell,-k,i}
    \end{pmatrix}.
\end{equation}

In this notation, the Green's functions of the central conductor become matrices in electron-hole space. For instance, the retarded Green's function of the conductor is defined as
\begin{equation}
    \bm{G}^{r}_{C}(t,t')
    := \bm{G}^{r}_{\bm{d}, \bm{d^\dagger}}(t,t')  \implies  \left(\bm{G}^{r}_{C}(t,t') \right)_{ij}
    = \left ( \bm{G}^{r}_{\bm{d}, \bm{d^\dagger}}(t,t') \right)_{ij}= \bm{G}^{r}_{{d}_i, d^\dagger_j}(t,t') =     -i\theta(t-t')
    \left\langle
    \left[
    {d}_i(t),
    d_j^{\dagger}(t')
    \right]_{+}
    \right\rangle,
    \label{Gdddagger}
\end{equation}
where the anticommutator $[A,B]_+ := AB + BA$ appears because the Nambu spinor is built from fermionic operators. Defining the outer product of a column vector $u$ of $n$ elements with a row vector $v$ of $m$ elements as $uv := M$ such that $M_{ij} := u_iv_j$, with $M$ being a $n \times m$ matrix, then we can write directly the definition in equation~\eqref{Gdddagger} in a compact notation as
\begin{equation}
    \bm{G}^{r}_{C}(t,t')
    := \bm{G}^{r}_{\bm{d}, \bm{d^\dagger}}(t,t') = -i\theta(t-t')
    \left\langle
    \left[
    \bm{d}(t),
    \bm{d}^{\dagger}(t')
    \right]_{+}
    \right\rangle,
\end{equation}

We can compute this Green's function using the Heisenberg equation of motion, which states that, given an operator $O$, then
\begin{equation}
    \label{eq:heisenberg}
    \begin{gathered}
        \frac{d}{dt} O_{\mathrm{H}}(t)
        =
        \frac{i}{\hbar}
        \left[
        H_{\mathrm{H}}(t), O_{\mathrm{H}}(t)
        \right]
        +
        \left(
        \frac{\partial O_{\mathrm{S}}}{\partial t}
        \right)_{\mathrm{H}},
        \\[0.4cm]
        \frac{d}{dt}
        \left\langle
        O_{\mathrm{H}}(t)
        \right\rangle
        =
        \frac{i}{\hbar}
        \left\langle
        \left[
        H_{\mathrm{H}}(t), O_{\mathrm{H}}(t)
        \right]
        \right\rangle
        +
        \left\langle
        \left(
        \frac{\partial O_{\mathrm{S}}}{\partial t}
        \right)_{\mathrm{H}}
        \right\rangle,
    \end{gathered}
\end{equation}
where $H_H$ is the Hamiltonian in Heisenberg picture, $O_H$ is the operator in Heisenberg picture and $O_S$ in Schrödinger picture, and $\left [A,B \right] := \left[A,B \right ]_- = AB - BA$ is the commutator. For the retarded Green's function $\bm{G}_{C}^r$, $H$ must be the the total Hamiltonian of the complete system (equation~\eqref{hamiltoniancomplet}). We find then that this Green's function can be expressed in terms of the Green's function of the isolated conductor $\bm g^{r}_{C}$, with a Hamiltonian $H_C$ only, as~\cite{Datta1995,HaugJauho2008}
\begin{equation}
    \bm G_{C}^{r}(\omega) = \left[ \bigl(\bm g_{C}^{r}(\omega)\bigr)^{-1} - \bm \Sigma_{N}^{r}(\omega) - \bm \Sigma_{S}^{r}(\omega) \right]^{-1},
    \label{green function }
\end{equation}
where the $\bm{\Sigma}_i^r$ are the retarded self-energies, defined as \begin{equation}
    \bm{\Sigma}_{i}^{r}(\omega)
    :=
    \bm{t}_{{C}i}\,
    \bm{g}_{i}^{r}(\omega)\,
    \bm{t}_{i{C}} ,
    \label{app:retarded-self-energy}
\end{equation}
with $\bm{t}_{Ci}$ is the coupling matrix between the conductor and lead $i$, whose elements are the coupling coefficients $t_{lki}$ in $H_T$. In general, the Green's functions associated only with isolated Hamiltonian $H_i$ are denoted by a lowercase $\bm{g}_i$, while the uppercase $\bm G$ is used for the Green's function of the complete Hamiltonian $H = \sum_i H_i$. The advanced self-energies are defined as \begin{equation}
    \bm{\Sigma}_{i}^{a}(\omega)
    :=
    \bm{t}_{{C}i}\,
    \bm{g}_{i}^{a}(\omega)\,
    \bm{t}_{i{C}} ,
    \label{app:advanced-self-energy}
\end{equation}
so we have that $\bm{\Sigma}_{i}^{a}(\omega)=[\bm{\Sigma}_{i}^{r}(\omega)]^{\dagger}$. Using the equation~\eqref{eq:heisenberg}, the Green's function $\bm{g}^r_C$ can be obtained from the corresponding commutation and anticommutation relations, yielding
\begin{equation}
    \left[
    \mathbb{I} \, i\frac{\partial}{\partial t}
    -
    \bm{H}^{\mathrm{BdG}}_C
    \right]
    \bm{g}^{r}_{C}(t,t')
    =
    \delta(t-t')\mathbb{I}.
    \label{greenisolateddifequation}
\end{equation}

In frequency space, the equation~\eqref{greenisolateddifequation} becomes
\begin{equation}
    \bm g^{r}_{C}(\omega) = \left[ (\omega + i0^{+}) \mathbb{I} - \mathbf{H}^{\mathrm{BdG}}_{C} \right]^{-1} .
    \label{eq:g lowercase}
\end{equation}
with $0^+ : = \lim\limits_{\eta \to 0^+} \eta$. Then, in frequency space, for a non-interacting conductor coupled to the two leads, the retarded Green's function can be written as
\begin{equation}
    \bm{G}^{r}_{C}(\omega)
    =
    \left[
    (\omega+i0^{+})\mathbb{I}
    -
    \bm{H}^{\mathrm{BdG}}_C
    -
    \bm{\Sigma}^{r}_{N}(\omega)
    -
    \bm{\Sigma}^{r}_{S}(\omega)
    \right]^{-1},
\end{equation}
We define the broadening matrix~\cite{Datta1995,HaugJauho2008} as
\begin{equation}
    \bm{\Gamma}_{i}(\epsilon)
    :=
    i\left[
    \bm{\Sigma}_{i}^{r}(\epsilon)
    -\bm{\Sigma}_{i}^{a}(\epsilon)
    \right] = i\left[
    \bm{\Sigma}_{i}^{r}(\epsilon)
    -\left [\bm{\Sigma}_{i}^{r}(\epsilon)
    \right]^\dagger \right] .
    \label{eq:general_broadening_matrix}
\end{equation}
The matrix $\bm{\Gamma}_{i}$ measures the spectral escape rate from the central region into lead $i$. For a superconducting reservoir, the self-energy is a Nambu matrix and can contain diagonal normal terms and off-diagonal anomalous terms. Below the ideal BCS gap, the quasiparticle broadening associated with propagating superconducting states vanishes, but the real anomalous part of $\bm{\Sigma}_{S}^{r}$ can still mix electron and hole sectors. This electron--hole mixing is precisely what allows Andreev reflection in the scattering matrix. Above the gap, $\bm{\Gamma}_{S}(\epsilon)$ is nonzero and describes escape into propagating superconducting quasiparticles.

The generalized Fisher--Lee relation~\cite{Fisher1981,Datta1995} expresses the scattering matrix in terms of $\bm{G}^{r}_C$. Firstly, it introduces coupling matrices $\bm{W}_{i}(\epsilon)$ satisfying
\begin{equation}
    \bm{W}_{i}(\epsilon)\bm{W}_{i}^{\dagger}(\epsilon)
    =
    \bm{\Gamma}_{i}(\epsilon) .
    \label{eq:general_W_square_root}
\end{equation}
The choice of $\bm{W}_{i}$ is not unique; any matrix square root of $\bm{\Gamma}_{i}$ gives the same physical scattering probabilities, up to a unitary rotation of the asymptotic lead-channel basis. The Fisher--Lee relation then reads
\begin{equation}
    \bm{s}_{ij}(\epsilon)
    =
    \delta_{ij}\,\mathbb{I}
    -i\,\bm{W}_{i}^{\dagger}(\epsilon)\,
    \bm{G}^{r}(\epsilon)\,
    \bm{W}_{j}(\epsilon),
    \label{eq:general_fisher_lee}
\end{equation}
where $\bm{s}_{ij}$ is the block of the full scattering matrix that connects outgoing electron and hole channels in lead $j$ with incoming electron and hole channels in lead $i$. Each block is itself a matrix in Nambu electron--hole space and, if present, in transverse channel space. In components,
\begin{equation}
    s_{ij}^{\alpha\beta}(\epsilon)
    =
    \delta_{ij}\delta_{\alpha\beta}
    -i\left[
    \bm{W}_{i}^{\dagger}(\epsilon)
    \bm{G}^{r}(\epsilon)
    \bm{W}_{j}(\epsilon)
    \right]_{\alpha\beta},
    \qquad
    \alpha,\beta\in\{\mathrm{e},\mathrm{h}\} .
    \label{eq:general_fisher_lee_components}
\end{equation}

\subsection{Green's functions and Fisher Lee relation in a Quantum Dot Normal--Superconducting structure}
We consider a single-level quantum dot coupled to a normal lead $N$ and to a superconducting lead $S$~\cite{deFranceschi2010,FazioRaimondi1998,Governale2008NS}. The dot is described by a resonant level Hamiltonian
\begin{equation}
    H_D
    =
    \sum_{\sigma}
    \varepsilon_d \,
    d^{\dagger}_{\sigma}
    d_{\sigma},
\end{equation}
where $\varepsilon_d$ is the dot level energy, tunable by a gate voltage, and $d_\sigma^\dagger, \,d_\sigma$ are fermionic operators that creates/annihilates an electron inside the dot. The fermionic commutative relations are
\begin{equation}
    \left \{d_i,d_j \right\} = \left\{d_i^\dagger, d_j^\dagger \right\} = 0, \qquad \left\{d_i, d_j^\dagger \right\} = \delta_{ij}
\end{equation}
with $\left \{A,B \right\} := \left [A,B \right ]_+ = AB+BA$, and $\delta_{ij}$ is the Kronecker delta. In the following we neglect Coulomb interactions and work in the Nambu basis
\begin{equation}
    \bm{d_\uparrow}
    :=
    \begin{pmatrix}
        d_{\uparrow}\\
        d^{\dagger}_{\downarrow}
    \end{pmatrix},
    \qquad
    \bm{d}^{\dagger}_\uparrow
    =
    \begin{pmatrix}
        d^{\dagger}_{\uparrow} &
        d_{\downarrow}
    \end{pmatrix}, \qquad     \bm{d_\downarrow}
    :=
    \begin{pmatrix}
        d_{\downarrow}\\
        d^{\dagger}_{\uparrow}
    \end{pmatrix},
    \qquad
    \bm{d}^{\dagger}_\downarrow
    =
    \begin{pmatrix}
        d^{\dagger}_{\downarrow} &
        d_{\uparrow}
    \end{pmatrix}.
\end{equation}
In this basis, the dot Hamiltonian is represented by
\begin{equation}
    H_D
    = \frac{1}{2} \sum_{\sigma}
    \bm{d}^{\dagger}_{\sigma}
    \bm{H}_C^{\mathrm{BdG}}
    \bm{d}_{\sigma} =
    \bm{d}^{\dagger}_{\uparrow}
    \bm{H}_C^{\mathrm{BdG}}
    \bm{d}_{\uparrow}, \qquad \text{with } \bm{H}^{\mathrm{BdG}}_D
    =
    \varepsilon_d \tau_z
    =
    \begin{pmatrix}
        \varepsilon_d & 0\\
        0 & -\varepsilon_d
    \end{pmatrix},
\end{equation}
being $\tau_z$ is the Pauli matrix acting in Nambu space. Here, the conductor is labeled with $D$ as a short of dot, instead of $C$ as a short of conductor in general. For a stationary system, the Green's functions only depend on the time difference and can be written in frequency space. The retarded Green's function is obtained from the Dyson equation~\cite{HaugJauho2008,PhysRevB.91.104518,PhysRevB.94.054506}
\begin{equation}
    \bm{G}^{r}_{D}(\epsilon)
    =
    \left[
    (\epsilon +i0^{+})\mathbb{I}
    -
    \varepsilon_d \tau_z
    -
    \bm{\Sigma}^{r}_{N}(\epsilon)
    -
    \bm{\Sigma}^{r}_{S}(\epsilon)
    \right]^{-1}.
    \label{eq:QD_Dyson_NS}
\end{equation}
Here $\bm{\Sigma}^{r}_{N}$ and $\bm{\Sigma}^{r}_{S}$ are the retarded self-energies induced by the normal and superconducting leads. In the wide-band limit, the normal lead self-energy is
\begin{equation}
    \bm{\Sigma}^{r}_{N}(\epsilon)
    =
    -\frac{i\Gamma_N}{2}
    \mathbb{I}
    =
    -\frac{i\Gamma_N}{2}
    \begin{pmatrix}
        1 & 0\\
        0 & 1
    \end{pmatrix},
    \label{eq:normal_self_energy}
\end{equation}
where $\Gamma_N$ is the tunnel coupling to the normal lead, defined later in equation~\eqref{eq: Gamma2}. The superconducting lead is described by a BCS self-energy~\cite{Blonder1982,Beenakker1997,PhysRevB.91.104518,PhysRevB.94.054506}. Choosing the superconducting order parameter real, $\Delta \in \mathbb{R}$, one has
\begin{equation}
    \bm{\Sigma}^{r}_{S}(\epsilon)
    =
    -\frac{\Gamma_S}{2}
    \frac{1}{\sqrt{\Delta^2-(\epsilon+i0^{+})^2}}
    \begin{pmatrix}
        \epsilon & \Delta\\
        \Delta^* & \epsilon
    \end{pmatrix} = -\frac{\Gamma_S}{2}
    \frac{1}{\sqrt{\Delta^2-(\epsilon+i0^{+})^2}}
    \begin{pmatrix}
        \epsilon & \Delta\\
        \Delta & \epsilon
    \end{pmatrix}.
    \label{eq:superconducting_self_energy_compact}
\end{equation}
Equivalently,
\begin{equation}
    \bm{\Sigma}^{r}_{S}(\epsilon)
    =
    \begin{cases}
        \displaystyle
        -\frac{\Gamma_S}{2\sqrt{\Delta^2-\epsilon^2}}
        \begin{pmatrix}
            \epsilon & \Delta\\
            \Delta & \epsilon
        \end{pmatrix},
        & |\epsilon|<\Delta,
        \\[0.5cm]
        \displaystyle
        -i\frac{\Gamma_S}{2\sqrt{\epsilon^2-\Delta^2}}
        \begin{pmatrix}
            |\epsilon| & \Delta\\
            \Delta & |\epsilon|
        \end{pmatrix},
        & |\epsilon|>\Delta.
    \end{cases}
    \label{eq:superconducting_self_energy_piecewise}
\end{equation}
where $\Gamma_S$ is the tunnel coupling to the superconducting lead, also defined later in equation~\eqref{eq: Gamma2}. The subgap self-energy is real, reflecting the absence of quasiparticle states in the superconductor for $|E|<\Delta$, while the above-gap self-energy acquires an imaginary part associated with quasiparticle tunneling.

The tunnel couplings are defined as
\begin{equation}
    \Gamma_{i\sigma}(\epsilon)
    :=
    2\pi
    \sum_k
    \left|t_{ik\sigma}\right|^2
    \rho_{ik\sigma}
    \label{eq:Gamma}
\end{equation}
where $\rho_{i k \sigma}$ is the density of states~\cite{Datta1995,HaugJauho2008}. For spin-independent tunneling, $t_{ik \sigma} = t_{ik}$ and $\rho_{ik\sigma} = \rho_{ik}$ $\forall \sigma$, so $\Gamma_{i\sigma}(\epsilon) = \Gamma_{i}(\epsilon)$  $\forall \sigma$. Furthermore, in the wide-band limit, $t_{ik} = t_i$  $\forall k$, so equation~\eqref{eq:Gamma} becomes
\begin{equation}
    \Gamma_{i\sigma} = \Gamma_i
    =
    2\pi
    \rho_i
    |t_i|^2,
    \label{eq: Gamma2}
\end{equation}
with $\rho_i = \sum_k \rho_{ik}$. Moreover, in this wide-band limit, $\Gamma_i$ is constant, i.e., it doesn't depend on the energy. The density of states of a quantum dot is
\begin{equation}
    \rho_{ik\sigma}(\epsilon)
    =
    \delta\!\left(\epsilon-\epsilon_{ik\sigma}\right),
\end{equation}
where $\delta(\xi)$ is the Dirac delta. The coupling matrices $\bm{\Gamma}_i$ entering the generalized Fisher--Lee relation~\cite{Fisher1981,Datta1995} are then
\begin{equation}
    \begin{gathered}
        \bm{\Gamma}_{N}(\epsilon)
        =
        \Gamma_N \mathbb{I}
        =
        \Gamma_N 
        \begin{pmatrix}
            1 & 0\\
            0 & 1
        \end{pmatrix},
        \\[0.4cm]
        \bm{\Gamma}_{S}(\epsilon)
        =
        \frac{\Gamma_S}{\sqrt{\epsilon^2-\Delta^2}}
        \begin{pmatrix}
            |\epsilon| & \Delta\\
            \Delta & |\epsilon|
        \end{pmatrix}
        \theta \left (|\epsilon|-\Delta \right).
    \end{gathered}
\end{equation}
Equation~\eqref{eq:QD_Dyson_NS} gives
\begin{equation}
    \bm{G}^{r}_{D}(\epsilon)
    =
    \frac{1}{D(\epsilon)}
    \begin{pmatrix}
        \epsilon+\varepsilon_d+\frac{i\Gamma_N}{2}-\Sigma^{r}_{S,22}(\epsilon)
        &
        \Sigma^{r}_{S,12}(\epsilon)
        \\
        \Sigma^{r}_{S,21}(\epsilon)
        &
        \epsilon-\varepsilon_d+\frac{i\Gamma_N}{2}-\Sigma^{r}_{S,11}(\epsilon)
    \end{pmatrix},
    \label{eq:QD_retarded_GF_explicit}
\end{equation}
with
\begin{equation}
        D(\epsilon)
        :=
        \left[
        \epsilon-\varepsilon_d+\frac{i\Gamma_N}{2}
        -
        \Sigma^{r}_{S,11}(\epsilon)
        \right]
        \left[
        \epsilon+\varepsilon_d+\frac{i\Gamma_N}{2}
        -
        \Sigma^{r}_{S,22}(\epsilon)
        \right]
         - \Sigma^{r}_{S,12}(\epsilon)
        \Sigma^{r}_{S,21}(\epsilon).
    \label{eq:QD_retarded_GF_determinant}
\end{equation}

This retarded Green's function is the only dot Green's function needed to construct the scattering matrix through the generalized Fisher--Lee relation. Due to microreversibility and particle-hole symmetry, it satisfies
\begin{equation}
    G^{r}_{D, 22}(\epsilon)=-\left[G^{r}_{D, 11}(-\epsilon)\right]^*,
    \qquad
    G^{r}_{D, 12}(\epsilon)=G^{r}_{D, 21}(-\epsilon),
        \qquad
    G^{r}_{D, 12}(\epsilon)=G^{r}_{D, 21}(\epsilon).
\end{equation}

With these quantities, the Andreev and quasiparticle transmission defined in equations~\eqref{andreevtransmission} and~\eqref{quasiparticletranmission} for a quantum dot are given by
\begin{equation}
    \begin{gathered}
        T_A(\epsilon)
        =
        \Gamma_N^2
        \left|
        G^{r}_{D,12}(\epsilon)
        \right|^2,
        \\[0.4cm]
        T_Q(\epsilon)
        = \begin{cases}
            0 & \text{if } |\epsilon| < \Delta, \\
        \Gamma_N
        \frac{\Gamma_S |\epsilon|}{\sqrt{\epsilon^2-\Delta^2}}
        \theta(|\epsilon|-\Delta)
        \left[
            \left|G^{r}_{D, 11}(\epsilon)\right|^2
            +
            \left|G^{r}_{D, 12}(\epsilon)\right|^2
            -
            \frac{2\Delta}{|\epsilon|}
            \operatorname{Re}
            \left(
            G^{r}_{D, 11}(\epsilon)
            \left[
            G^{r}_{D, 12}(\epsilon)
            \right]^*
            \right)
        \right] & \text{if }|\epsilon| > \Delta .
        \end{cases}
    \end{gathered}
\end{equation}

We note that, with the definitions introduced above and using the generalized Fisher--Lee relation between the scattering amplitudes and the Green's functions, the charge current $I$, the energy current $J^E$, the heat currents $J_i$, and the Andreev and quasiparticle transmissions $T_A$ and $T_Q$ reproduce, for an N--QD--S realization, the standard expressions obtained via alternative methods, such as directly applying the Heisenberg equation of motion to the current and energy operators within the nonequilibrium Green's function formalism~\cite{PhysRevB.91.104518,PhysRevB.94.054506}.

\clearpage
% Single, unified bibliography for the main article and all appendices.
\bibliographystyle{plainnat}
\bibliography{bibliography_arxiv}

\end{document}